\documentclass{emulateapj}

\usepackage{natbib}
\usepackage{apjfonts}
\begin{document}

\title{A Stellar Dynamical Mass Measurement of the Black Hole in NGC
  3998 from Keck Adaptive Optics Observations}

\author{Jonelle L. Walsh$^{1,2}$, Remco C.~E. van den Bosch$^3$,
  Aaron J. Barth$^2$, {\sc and} Marc Sarzi$^4$}

\affil{$^1$Department of Astronomy, The University of Texas at Austin,
  1 University Station C1400, Austin, TX 78712, USA;
  jlwalsh@astro.as.utexas.edu\\
  $^2$Department of Physics and Astronomy, University of California at
  Irvine, 4129 Frederick Reines Hall, Irvine, CA 92697, USA\\
  $^3$Max-Planck Institut f\"{u}r Astronomie, K\"{o}nigstuhl 17,
  D-69117 Heidelberg, Germany\\
  $^4$Centre for Astrophysics Research, University of Hertfordshire,
  AL10 9AB Hatfield, UK}

\begin{abstract}

  We present a new stellar dynamical mass measurement of the black
  hole in the nearby, S0 galaxy NGC 3998. By combining laser guide
  star adaptive optics observations obtained with the OH-Suppressing
  Infrared Imaging Spectrograph on the Keck II telescope with
  long-slit spectroscopy from the \emph{Hubble Space Telescope} and
  the Keck I telescope, we map out the stellar kinematics on both
  small spatial scales, well within the black hole sphere of
  influence, and on large scales. We find that the galaxy is rapidly
  rotating and exhibits a sharp central peak in the velocity
  dispersion. Using the kinematics and the stellar luminosity density
  derived from imaging observations, we construct three-integral,
  orbit-based, triaxial stellar dynamical models. We find the black
  hole has a mass of $M_\mathrm{BH} = (8.1_{-1.9}^{+2.0}) \times 10^8\
  M_\odot$, with an $I$-band stellar mass-to-light ratio of $M/L =
  5.0_{-0.4}^{+0.3}\ M_\odot/L_\odot$ ($3\sigma$ uncertainties), and
  that the intrinsic shape of the galaxy is very round, but
  oblate. With the work presented here, NGC 3998 is now one of a very
  small number of galaxies for which both stellar and gas dynamical
  modeling have been used to measure the mass of the black hole. The
  stellar dynamical mass is nearly a factor of four larger than the
  previous gas dynamical black hole mass measurement. Given that this
  cross-check has so far only been attempted on a few galaxies with
  mixed results, carrying out similar studies in other objects is
  essential for quantifying the magnitude and distribution of the
  cosmic scatter in the black hole mass $-$ host galaxy relations.

\end{abstract}


\section{Introduction}
\label{sec:intro}

Black holes reside at the centers of most massive galaxies
\citep{Magorrian_1998, Richstone_1998} and their masses
($M_\mathrm{BH}$) are known to tightly correlate with properties of
their host galaxy, like the bulge stellar velocity dispersion
($\sigma_\star$; e.g., \citealt{Ferrarese_2000, Gebhardt_2000b,
  Tremaine_2002, Gultekin_2009b, Graham_2011}) and the bulge
luminosity ($L_\mathrm{bul}$; e.g., \citealt{Dressler_1989,
  Kormendy_Richstone_1995, Marconi_Hunt_2003,
  Gultekin_2009b}). Playing a fundamental role in establishing the
empirical $M_\mathrm{BH}$ $-$ host galaxy relationships over the past
15 years has been the \emph{Hubble Space Telescope} (\emph{HST}). In
fact, \emph{HST} spectroscopic observations have lead to about 50 of
the 70 black hole mass measurements made to date.

More recently, adaptive optics (AO) systems on large ground-based
telescopes have become an attractive alternative for obtaining high
spatial resolution data. These systems correct for wavefront
distortions caused by turbulence in the Earth's atmosphere using a
natural guide star or a laser guide star (LGS) as a reference. They
are often used in conjunction with an integral field unit (IFU), which
can map out the two-dimensional (2D) velocity field very efficiently
compared to traditional long-slit spectrographs. In addition, AO
systems operating in the near infrared on $8-10$m telescopes are able
to probe deep into the central regions of galaxies that previously
presented significant observational challenges with the 2.4m
\emph{HST}, such as giant ellipticals with low-surface brightness
cores and galaxy nuclei obscured by dust. AO-assisted IFUs will
undoubtedly have a major impact on the field of supermassive black
hole detection, and black hole mass measurements using this technology
have already become increasingly more prevalent in the literature
\citep[e.g.,][]{Houghton_2006, Davies_2006, Neumayer_2007, Nowak_2010,
  Rusli_2011}.

Measurements of black hole masses are most often made through the
dynamical modeling of gas disks or stars. Gas dynamical modeling is
conceptually simple, and the method can be applied to galaxies
containing nuclear gas disks in circular rotation. The main drawback
of this technique is that the gas can be influenced by
non-gravitational forces, and the assumption of circular rotation must
be verified. Furthermore, many of the observed gas disks in early-type
galaxies exhibit velocity dispersions in excess of those expected from
just rotational motion alone. This internal velocity dispersion can be
quite large, ranging from $\sim$$100 - 500$ km s$^{-1}$
\citep[e.g.,][]{vanderMarel_vandenBosch_1998, VerdoesKleijn_2002,
  DallaBonta_2009}. The physical nature of the intrinsic velocity
dispersion is not understood. One interpretation is that the intrinsic
velocity dispersion is the result of local microturbulence, but the
bulk motion of the gas remains at the circular velocity
\citep{vanderMarel_vandenBosch_1998}. Alternatively, the intrinsic
velocity dispersion could be dynamically important, providing pressure
support to the gas disk. Models that do not account for this effect
will underestimate the true black hole mass
\citep[e.g.,][]{Barth_2001, Neumayer_2007, Walsh_2010}.

In contrast, stellar dynamical modeling is the most widely used
method. However, extracting a secure black hole mass measurement from
the observed kinematics can be challenging due to the complexity of
the models. Recent adjustments to several previous stellar dynamical
measurements have been made after identifying significant systematic
effects due to degeneracies between the dark matter halo and stellar
mass-to-light ratio \citep{Gebhardt_Thomas_2009}, insufficient
coverage of the phase space occupied by tangential orbits
\citep{Shen_Gebhardt_2010, Schulze_Gebhardt_2011}, and the effects of
triaxiality \citep{vandenBosch_deZeeuw_2010}. Other possible sources
of error, such as the choice of regularization \citep{Valluri_2004}
and the uncertain inclinations of early-type galaxies
\citep[e.g.,][]{Verolme_2002, Gebhardt_2003, Krajnovic_2005}, have
lead to further debate about the accuracy of stellar dynamical mass
measurements.

Since gas and stellar dynamical modeling methods suffer from different
systematic effects, carrying out consistency tests between the two
techniques is crucial. Only by applying both mass measurement methods
to the same object can we address important unanswered questions
regarding the black hole scaling relations. Is there a systematic
difference between the masses derived from the two methods, and if so,
how does this affect the slope of the relations? How much of the
scatter in the $M_\mathrm{BH}$ $-$ host galaxy relationships is the
result of inconsistencies between stellar and gas dynamical
measurements?  Theoretical interpretations of $M_\mathrm{BH}$ $-$ host
galaxy correlations range from the black hole being an essential
component in galaxy evolution \citep[e.g.,][]{Silk_Rees_1998,
  DiMatteo_2005} to simply being the result of random mergers without
the need for a coevolution of black holes and galaxies
\citep{Peng_2007, Jahnke_Maccio_2011}. Understanding the shape and
cosmic scatter of the relations, especially at the sparsely populated
upper- and lower-mass ends, are key to distinguishing between the
various theories and for characterizing the black hole mass function.

This necessary consistency check has only been attempted on a few
galaxies with mixed results. Stellar and gas dynamical models have
been applied to IC 1459 \citep{VerdoesKleijn_2000, Cappellari_2002b}
and NGC 3379 \citep{Gebhardt_2000a, Shapiro_2006,
  vandenBosch_deZeeuw_2010}. Unfortunately, in both these galaxies,
the gas kinematics turned out to be disturbed and not compatible with
regular disklike rotation. Consequently, the ionized gas does not give
a useful measurement of the black hole mass, and the galaxies are not
suitable for a proper comparison between the two methods. In the cases
of NGC 3227 and NGC 4151, the gas dynamical black hole mass
measurements by \cite{Hick_Malkan_2008} are generally consistent with
the stellar dynamical masses reported by \cite{Davies_2006} and
\cite{Onken_2007}. However, \cite{Onken_2007} could not find a single
best fitting model for NGC 4151, and label their stellar dynamical
result as tentative. \cite{VerdoesKleijn_2002} applied both methods to
NGC 4335, and found a stellar dynamical mass that is at least five
times larger than the gas dynamical determination. The black holes in
M87 and Cen~A have been the subject of numerous gas and stellar
dynamical studies. While early examinations of the stellar dynamics
(see review by \citealt{Kormendy_Richstone_1995}, and references
therein) produced similar mass estimates for the black hole in M87 as
the gas dynamical measurements \citep{Harms_1994, Macchetto_1997}, the
most recent stellar dynamical mass from \cite{Gebhardt_2011} is about
a factor of two larger. Finally, there is a great variation in the
Cen~A gas dynamical black hole mass measurements, mainly due to
uncertainties in the inclination of the gas disk
(\citealt{Marconi_2001}, 2006, \citealt{HaringNeumayer_2006},
\citealt{Krajnovic_2007}, \citealt{Neumayer_2007}). The latest gas
dynamical measurement of \cite{Neumayer_2007} is significantly smaller
than the stellar dynamical measurement of \cite{Silge_2005} (by about
a factor of four), but is in agreement with the stellar dynamical
measurement by \cite{Cappellari_2009}. With only a few direct
comparisons between the gas and stellar dynamical techniques, clearly
more objects need to be examined.

An excellent target for such a comparison study is NGC 3998, which is
a nearby, S0 galaxy with a low ionization nuclear emission-line region
(LINER) nucleus \citep{Ho_1997}. The galaxy has a simple morphology
suitable for both stellar and gas dynamical modeling. The nuclear gas
kinematics have been previously studied using multiple slit positions
of the \emph{HST} Space Telescope Imaging Spectrograph (STIS), and
have been fit with a circularly rotating thin-disk model
\citep{deFrancesco_2006}. With the \cite{deFrancesco_2006} measurement
of $2.2 \times 10^8\ M_\odot$ (scaled to our assumed distance), the
black hole sphere of influence ($r_\mathrm{sphere} =
GM_\mathrm{BH}/\sigma_\star^2$) is expected to be 0\farcs15, which can
be resolved by IFUs combined with AO on 8$-$10m ground-based
telescopes. NGC 3998 also contains a bright, compact nucleus that is
suitable for use as a LGS tip-tilt reference. With the bulge stellar
velocity dispersion reported by \cite{Gultekin_2009b} of 305 km
s$^{-1}$, this galaxy is further interesting as falls at the upper end
of the $M_\mathrm{BH} - \sigma_\star$ relationship.

In this paper, we describe a measurement of the black hole in NGC 3998
using orbit-based stellar dynamical models. We will compare the
stellar dynamical black hole mass measurement to the existing gas
dynamical measurement by \cite{deFrancesco_2006} in order to test
whether the two methods give consistent results when applied to the
same galaxy. We begin by presenting the imaging observations and the
determination of the galaxy's stellar mass distribution in \S
\ref{sec:img_obs} and \S \ref{sec:stellarmass}. We describe the
spectroscopic observations in \S \ref{sec:spec_obs}, and the
measurement of the stellar kinematics in \S
\ref{sec:kinematics}. Models of the point-spread function (PSF) are
discussed in \S \ref{sec:psf}. In \S \ref{sec:modeling}, we provide a
brief overview of the stellar dynamical technique, and its application
to NGC 3998, as well as present the modeling results in \S
\ref{sec:results}. Finally, in \S \ref{sec:discussion}, we examine the
stellar orbital structure of the galaxy, compare the stellar dynamical
mass measurement to gas dynamical determination, and place the black
hole in NGC 3998 on the $M_{\mathrm{BH}}$ $-$ host galaxy
relationships. Throughout this paper, we adopt a distance to NGC 3998
of 13.7 Mpc \citep{Tonry_2001, Mei_2007}.

\section{Imaging Observations}
\label{sec:img_obs}

Imaging observations are essential components in the orbit-based
stellar dynamical models, and are used to derive the galaxy's
luminosity density. We obtained a Wide Field Planetary Camera 2
(WFPC2) F791W image of NGC 3998 centered on the Planetary Camera (PC)
detector from the \emph{HST} archive. The \emph{HST} image of NGC 3998
was originally acquired under program GO-5924, and the total exposure
time was 100 s. The WFPC2 image, with a pixel scale of 0\farcs046 and
0\farcs1 for the PC and Wide-Field chips, provides high angular
resolution imaging suitable for constructing a luminous mass model
near the black hole, however the image only extends out to a radius of
$\sim $10\arcsec\ (0.6 kpc). We therefore also included in the
analysis a deep $Ks$-band image (courtesy of L\"{a}sker et al., in
preparation) taken with the Wide-field InfraRed Camera
\citep[WIRCam;][]{Puget_2004} on the 3.6m Canada-France-Hawaii
Telescope (CFHT). The image has a spatial scale of 0\farcs3
pixel$^{-1}$ and covers a 30\arcmin$\times$30\arcmin\ field. The total
exposure time was 442 s, and we were able to measure the galaxy's
surface brightness profile out to a radius of $\sim $250\arcsec\ (16.6
kpc). The ground-based image was used to generate the mass model on
large spatial scales, which is helpful in constraining the intrinsic
shape of the galaxy in the stellar dynamical models.

\section{Stellar Mass Profile}
\label{sec:stellarmass}

\begin{figure}
\begin{center}
\epsscale{1.0}
\plotone{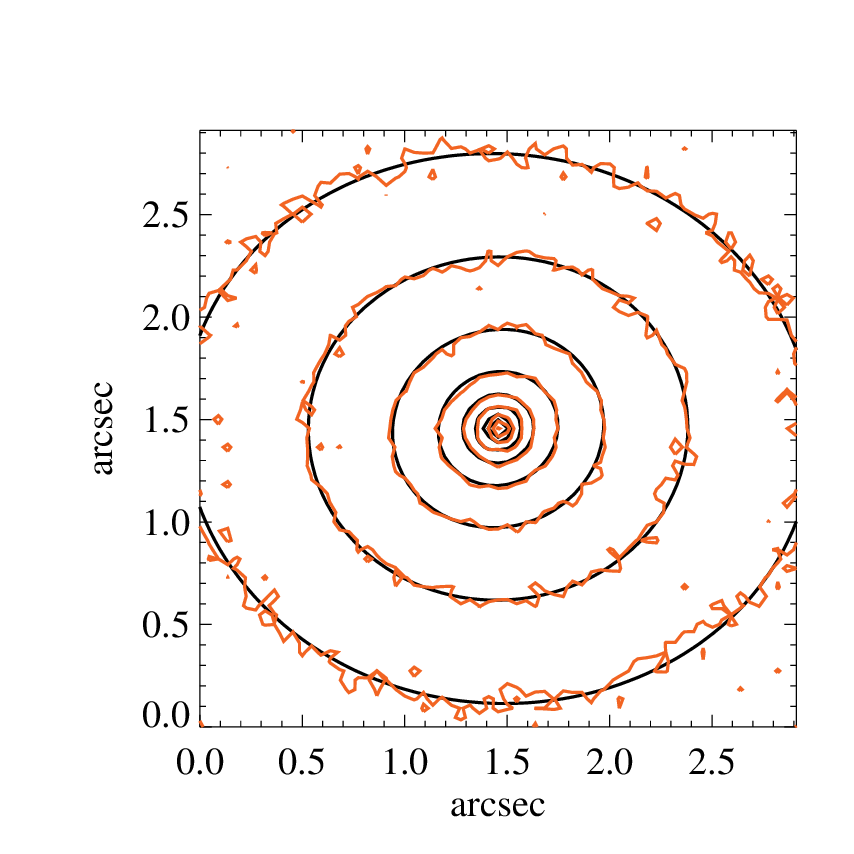} \\
\plotone{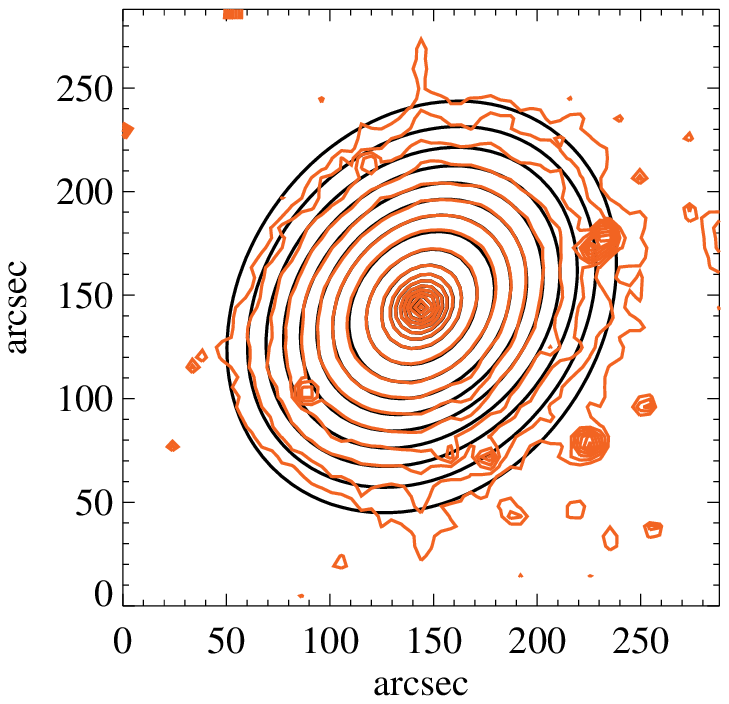}
\caption{Isophotes of the MGE model (black) are compared to the inner
  regions of the \emph{HST} WFPC2/PC F791W image (top) and the CFHT
  WIRCam $Ks$-band image (bottom) of NGC 3998. Contours are
  logarithmically spaced, but arbitrary. \label{fig:mge2d}}
\end{center}
\end{figure}

\begin{figure}
\begin{center}
\epsscale{0.9}
\plotone{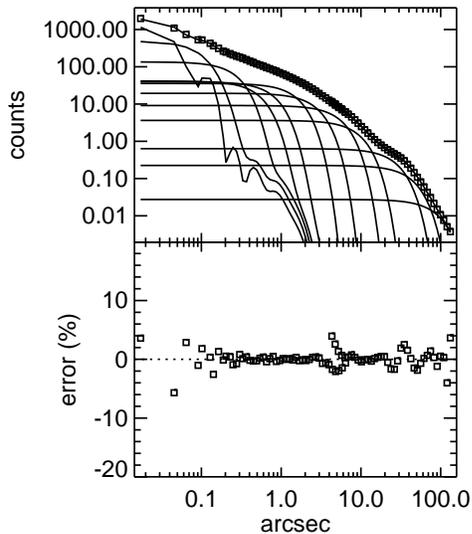}
\caption{\emph{Top panel:} Comparison between the observed surface
  brightness profile (open squares) and the best-fitting MGE model
  (solid line) averaged along the azimuthal axis. Individual
  components making up the MGE model, which have been convolved with
  the \emph{HST} PSF, are also shown. \emph{Bottom panel:} The percent
  error between the model and data. \label{fig:mge1d}}
\end{center}
\end{figure}

We parameterized the WFPC2 F791W and WIRCam $Ks$-band images of NGC
3998 as the sum of 2D Gaussians using the Multi-Gaussian Expansion
(MGE) method \citep{Cappellari_2002a}. The MGE method has often been
used to model high-resolution and ground-based images
\citep[e.g.,][]{Verolme_2002, Cappellari_2006, Onken_2007,
  DallaBonta_2009, Krajnovic_2009}, and we used the MGE method here
because it has the advantage of being able to reproduce a large range
of density distributions while also allowing for the deprojection to
be carried out analytically. MGE models were fit to both images
simultaneously while also accounting for the \emph{HST} PSF. The PSF
itself was described as the sum of 25 positive and negative Gaussians,
which were found by applying the MGE software to a Tiny Tim model
\citep{Krist_Hook_2004}. Our best-fit MGE parametrization of the
galaxy was composed of 12 Gaussians, where the innermost Gaussian was
constrained to be round and the position angles (PA) of the Gaussian
components were required to be the same. This model produced an
excellent fit to the imaging data, as can be seen in Figure
\ref{fig:mge2d}. In addition, in Figure \ref{fig:mge1d}, we present
the surface brightness profile, the MGE model, and the percent error
averaged along the azimuthal axis. The MGE model was corrected for
galactic extinction using the \cite{Schlegel_1998} values given by the
NASA/IPAC Extragalactic Database and the surface density was converted
to $I$-band solar units using the WFPC2 calibration by
\cite{Dolphin_2000}. In Table \ref{tab:mge}, we provide the best-fit
values of the MGE parameters. We further note that there are no
significant emission lines within the F791W bandpass \citep{Ho_1993}.

Although our adopted MGE model requires each Gaussian component to
have the same PA, we fit an additional MGE model that allowed for
isophotal twists. We detected very small changes to the PA, of
typically $<1^\circ$, between the components. Furthermore, when using
this MGE parameterization as input into the orbit-based stellar
dynamical models, we found worse agreement between the observed and
predicted stellar kinematics. As a result, we do not consider the MGE
model that allows for isophotal twists any longer, and focus solely on
the MGE parametrization that constrains each component to have the
same PA.

When constructing mass models of galaxies containing an active
galactic nucleus (AGN), often the innermost Gaussian of the MGE model
is assumed to arise from non-thermal emission and is excluded from the
stellar mass distribution. The nucleus of NGC 3998 has been
spectroscopically classified as a Type 1.9 LINER \citep{Ho_1997}, and
an unresolved, variable UV source, a compact radio source, and an
X-ray source have all been detected at the galaxy's center
\citep{Hummel_1984, Fabbiano_1994, Maoz_2005, Roberts_Warwick_2000,
  Pellegrini_2000}. All of this evidence suggests that NGC 3998 hosts
an AGN. However, the galaxy also exhibits a cuspy surface brightness
profile, and some starlight may still be contained within the
innermost Gaussian. We therefore ran orbit-based stellar dynamical
models using MGE parameterizations that both included and excluded
this central Gaussian, and will discuss the results in \S
\ref{sec:results} and \S \ref{subsec:more_errors}. These descriptions
represent the two extremes: in one case all the light from the
innermost Gaussian is attributed to the stars and in the other
scenario all the light is assigned to the AGN.

\begin{deluxetable}{lccc}
\tabletypesize{\scriptsize}
\tablewidth{0pt}
\tablecaption{Best-fit MGE Parameters \label{tab:mge}}
\tablehead{
\colhead{$j$} & 
\colhead{$\log\ I_j$ (L$_{\odot, I}$ pc$^{-2}$) } &
\colhead{$\sigma_j^\prime$ (\arcsec)} &
\colhead{$q_j^\prime$} \\
\colhead{(1)} &
\colhead{(2)} &
\colhead{(3)} &
\colhead{(4)}
}

\startdata

1   &   6.445  &   0.017  &   0.990 \\
2   &   5.405  &   0.078  &   0.940 \\
3   &   4.808  &   0.223  &   0.962 \\
4   &   4.445  &   0.504  &   0.946 \\
5   &   3.814  &   1.037  &   0.756 \\
6   &   4.096  &   1.200  &   0.903 \\
7   &   3.973  &   2.042  &   0.881 \\
8   &   3.641  &   4.134  &   0.799 \\
9   &   3.347  &   7.223  &   0.813 \\
10  &   2.444  &  21.210  &   0.796 \\
11  &   2.149  &  34.812  &   0.776 \\
12  &   1.034  &  73.492  &   0.866

\enddata

\tablecomments{Column (1) gives the Gaussian component number, column
  (2) is central surface brightness of each Gaussian, column (3)
  provides the dispersion of Gaussian along the major axis, and column
  (4) lists the axis ratio. All of the Gaussian components were
  constrained to have the same position angle, which was determined to
  be $-41.5^\circ$. The primed variables indicate projected
  quantities. Stellar dynamical models were run using MGE models that
  both include and exclude the central component ($j = 1$).}

\end{deluxetable}

\section{Spectroscopic Observations}
\label{sec:spec_obs}

We observed NGC 3998 using the IFU OH-Suppressing Infrared Imaging
Spectrograph \citep[OSIRIS;][]{Larkin_2006} assisted by the LGS AO
system \citep{Wizinowich_2006, vanDam_2006} on the 10-m Keck II
telescope. The OSIRIS data provided high angular resolution
observations of the stellar kinematics over a 2D field and allowed us
to resolve $r_\mathrm{sphere}$. We also retrieved archival STIS
observations of NGC 3998 from the \emph{HST} archive. The STIS
observations have a well-characterized PSF and provided us with
additional high-resolution data of the stellar kinematics along a
single PA. Preliminary work from a stellar dynamical analysis using
this same STIS data was described by \cite{Bower_2000}. Past studies
have shown the importance of acquiring large-scale observations of the
kinematics in order to properly constrain the orbital distribution in
the stellar dynamical models \citep[e.g.,][]{Shapiro_2006,
  Krajnovic_2009}. For this reason, we also obtained long-slit
observations using the Low-Resolution Imaging Spectrograph
\citep[LRIS;][]{Oke_1995} on the 10-m Keck I telescope. In the
sections that follow we describe the OSIRIS, STIS, and LRIS
observations and data reduction.

\subsection{LGS AO OSIRIS Observations}
\label{subsec:osiris_obs}

The OSIRIS observations were acquired on 2009 May 1 and 2 during the
first half of each night. We used the 0\farcs05 spatial scale and the
broadband $K$ (Kbb) filter, providing coverage of $1.965 - 2.381\
\mu$m. We aligned the long axis of the IFU with the major axis of the
nuclear emission-line disk at PA = $308^\circ$
\citep{deFrancesco_2006}. Using the nucleus of NGC 3998 as the
tip-tilt reference, we were able to achieve a good AO correction and
we estimate that the PSF core had a full width at maximum intensity
(FWHM) of $\sim$0\farcs1 (see \S \ref{sec:psf}). We obtained 300 s
exposures of the galaxy nucleus and sky, following the sequence
Object-Object-Object-Sky Object-Object-Object-Sky, and recorded 3.8
hours of on-source integration time.  Each object exposure was
dithered 0\farcs2 perpendicular to the long-axis of the IFU in order
to remove bad pixels and to obtain a slightly wider field of view of
1\farcs2 $\times$ 3\farcs2. In addition, we observed K and M giant
stars for use as velocity templates in the same observational setup as
was used for the galaxy, and A0~V stars for telluric correction.

The data were reduced using the OSIRIS data reduction pipeline
v2.3\footnote[5]{http://irlab.astro.ucla.edu/osiris/pipeline.html}. The
pipeline includes subtraction of a sky frame, glitch identification
and cosmic ray cleaning, extraction of the spectra into a data cube
(with two spatial dimensions, $x$ and $y$ corresponding to the short
and long axis of the IFU, and one spectral dimension), wavelength
calibration, atmospheric dispersion correction, and telluric
correction using the extracted one-dimensional (1D) spectrum of an
A0~V star. While the reduction pipeline is capable of performing
scaled sky subtraction following the algorithm outlined by
\cite{Davies_2007}, we found that the current implementation did not
work well for our data set and produced strong sky residuals in the
reduced cubes. We instead opted for a direct subtraction method, where
we subtracted a sky exposure from the previous two object frames and
from the following object frame. Thus, the sky and object frames were
separated in time by at most 600 s, and the direct subtraction method
worked well. After all of the individual galaxy cubes were reduced, we
determined the spatial $x$ and $y$ offsets between the first galaxy
cube and each of the remaining object cubes. This was done by summing
along the wavelength axis of the data cubes to generate flux maps,
then cross-correlating the images. Using these offsets, the individual
object cubes were aligned and averaged to produce the final galaxy
data cube. The velocity template star observations were reduced in a
similar manner, but an additional step was included to extract a 1D
spectrum from the data cube using a circular aperture with a radius of
7 pixels.

\subsection{Archival STIS Observations}
\label{subsec:stis_obs}

We retrieved STIS observations of NGC 3998 made on 1999 March 10 under
program GO-7350 from the \emph{HST} archive. NGC 3998 was observed
with the {\tt 52x0.2} aperture at eight slit positions, all of which
were aligned at a PA of $152^\circ$, within $15^\circ$ of the galaxy's
major axis. Two successive exposures were obtained at each slit
position to aid in the removal of cosmic rays, and a shift of about
four pixels in the spatial direction was made between positions. The
average exposure time recorded at each slit position was 2730 s. The
G750M grating, centered on 8561 \AA, was read out in an unbinned mode,
which provided a scale of 0\farcs0507 pixel$^{-1}$ along the spatial
axis and 0\farcs554 \AA\ pixel$^{-1}$ along the dispersion
direction. Following the observation of the galaxy at each slit
position, a contemporaneous flat field was taken.

In addition to the galaxy observations, the K3 III star HR 260 was
observed as part of Program 7350 for use as a velocity template. In
order to build up a larger library, we also searched the \emph{HST}
archives for other suitable stars, using as a guide the previously
published stellar dynamical black hole mass measurements made from
STIS data. We retrieved an additional four stars, observed using the
same STIS setup, from the archive: the K0 III star HR 7615 (GO-7566),
the G8 III star HD 141680 (GO-8591), and the K2 III and G5 V stars HD
73471 and HD 115617 (GO-8928).

The data were reduced by running individual IRAF\footnote[6]{IRAF is
  distributed by the National Optical Astronomy Observatory, which is
  operated by the Association of Universities for Research in
  Astronomy under cooperative agreement with the National Science
  Foundation} tasks within the standard Space Telescope Science
Institute (STScI) CALSTIS pipeline. With these tasks, the overscan
region was trimmed, the bias and dark files were subtracted,
flat-field corrections were applied, CR-split exposures were combined
to reject cosmic rays, wavelength and flux calibration were performed,
and the data were rectified for geometric distortions. We chose to run
individual segments of the pipeline independently, following the
procedure detailed in the STIS Data Handbook v6.0
\citep{Bostroem_Proffitt_2011}, instead of running the wrapper CALSTIS
program so that a couple of important modifications could be made.

Specifically, we found that the dark calibration file provided by
STScI, which is constructed from several long dark exposures taken
over the course of a week, contained very strong hot pixels that did
not subtract well and left numerous negative pixels in the galaxy
spectral images. This poor subtraction is likely due to pixels that
vary on timescales of less than a week. We therefore followed a
similar approach to \cite{Bower_2001}, and modified the dark file by
replacing any pixels that deviated by more than $8\sigma$ with a
median value. After the initial processing was complete, we included
an additional cleaning step and used LA-COSMIC \citep{vanDokkum_2001}
to remove the hot pixels remaining after the subtraction our modified
dark file. We then applied the geometric distortion correction to the
cleaned images. Additionally, we used the contemporaneous flat field
exposures taken at each of the slit positions to remove the obvious
fringe pattern affecting the spectra. IRAF tasks were used to reduce
and normalize the flat field observations, shift and scale the fringes
in the normalized flat such that they would match those in the galaxy
images, and to finally apply the fringe correction.

The reduced, geometrically rectified, fringe-corrected spectral images
at each of the eight slit positions taken along the galaxy's major
axis were then aligned and combined to produce the final 2D galaxy
spectrum. The same procedure was applied to the five velocity standard
stars observed with STIS, but as a final step, we constructed a 1D
spectrum by adding together 3 spectral rows above and below the
central row from the 2D images.

\subsection{Long-Slit LRIS Observations}
\label{subsec:lris_obs}

The LRIS observations were obtained during the first half night on
2009 April 15. On the red side, we used the 831 lines mm$^{-1}$
grating centered on 8200 \AA\ with a 1\arcsec-wide slit, producing a
scale of 0.92 \AA\ pixel$^{-1}$ in the dispersion direction and
0\farcs211 pixel$^{-1}$ in the spatial direction. We placed the
long-slit along four PAs: along the major axis of the nuclear gas disk
\citep[PA $ = 308^\circ$;][]{deFrancesco_2006}, along the minor axis
(PA = 218$^\circ$), and at two intermediate angles (PA = 353$^\circ$
and PA = 263$^\circ$). At each PA we acquired a $2 \times 600$ s
observation, with the exception of the minor axis PA, where we
recorded a $2 \times 300$ s observation. Between individual exposures,
we dithered by 90\arcsec\ along the length of the slit. We
additionally observed K giant stars for use as velocity templates, and
the flux standard star Feige 34, in the same observational setup as
was employed for NGC 3998.

We reduced the LRIS data using IRAF. The initial processing steps of
the 2D spectral images included trim and bias subtraction, flat
fielding, and cosmic ray cleaning. For the NGC 3998 exposures, we also
geometrically rectified the 2D spectra so that both the wavelength and
spatial axes would run parallel along rows and columns,
respectively. During this process, we used the Hg, Ar, and Ne arc lamp
exposures acquired immediately following the NGC 3998 observations to
wavelength calibrate the spectral images. We removed the sky
background by subtracting the two dithered spectral images taken at
each PA from one another. The difference images were spatially
aligned, averaged together, and trimmed to remove the negative
features resulting from the pair subtraction. As a final reduction
step, we compared a calibrated 1D standard star spectrum to a 1D
spectrum of the galaxy extracted using the {\tt APALL} task to
determine the multiplicative factor (as a function of wavelength) that
was needed to bring the two spectra into agreement. The scaling was
then applied to the 2D galaxy spectrum to produce a final,
flux-calibrated spectral image of NGC 3998 at each PA.

The velocity template stars were reduced following a slightly
different procedure because we only required a 1D spectrum from the 2D
image. After the initial image processing, we applied the same
spectral rectification and wavelength calibration to the 2D spectra of
the velocity template stars as was used for the NGC 3998 spectra. We
performed a secondary correction to the wavelength calibration by
applying a small linear shift along the dispersion axis so that two
bright sky lines bounding the \ion{Ca}{2} triplet region matched the
known wavelengths of 8430 \AA\ and 8886 \AA. We then extracted a 1D
spectrum with {\tt APALL} using a 5\farcs3 aperture radius and sky
regions that were 8\farcs4 in width beginning at a distance of
23\arcsec. The velocity template spectra were flux calibrated using an
extracted 1D spectrum of the standard star Feige 34.

\section{Measuring the Stellar Kinematics}
\label{sec:kinematics}

From the OSIRIS, STIS, and LRIS data, we extracted 1D spectra over a
range of spatial locations and measured the line-of-sight velocity
distribution (LOSVD), which we parameterized with the first four
Gauss-Hermite moments ($V$, $\sigma$, $h_3$, and $h_4$). Measurement
of the LOSVD requires spectra with a high signal-to-noise ratio
(S/N). Typically, a S/N of $\sim$40 per spectral and spatial
resolution element are needed in order to reliably determine $h_3$ and
$h_4$, which quantify asymmetric and symmetric deviations from a
Gaussian \citep[e.g.,][]{vanderMarel_Franx_1993, Bender_1994,
  Statler_1995, Krajnovic_2009}. We used the Voronoi binning algorithm
\citep{Cappellari_Copin_2003} in order to achieve the best spatial
resolution possible given such a requirement on the minimum S/N. The
stellar kinematics in each of the spatial bins were then measured
using the penalized pixel fitting (pPXF) method of
\cite{Cappellari_Emsellem_2004}. With this method, logarithmically
rebinned galaxy spectra are fit in pixel space using a stellar
template that is convolved with the LOSVD.

The pPXF code minimizes template mismatch by finding an optimal
stellar template that is composed of a linear combination of spectra
from a library. The libraries were designed to contain stars that are
representative of the galaxy's expected stellar population. When
measuring the stellar kinematics from the OSIRIS data, we used a
template library composed of 20 stars, consisting of mostly K and M
giants (spectral types K0 $-$ K5 and M0 $-$ M5), a K4 V star, and a G9
V star. The STIS kinematics were determined using a library of five
stars (G8 III $-$ K3 III, and G5 V), and the LRIS kinematics were
extracted using a library of eight K0 III $-$ K3 III stars.

The optimal stellar template was determined just once by using pPXF to
fit a very high S/N spectrum generated by summing the spectra over all
spatial locations. Once the optimal stellar template was found, pPXF
was used to measure the kinematics in each spatial bin by holding
fixed the relative weights of the stars that comprise the optimal
template. We did, however, allow the coefficients of an additive
Legendre polynomial of degree 1 and a multiplicative Legendre
polynomial of degree 2 to vary. The additive polynomial was used to
model the AGN continuum, and the multiplicative polynomial was
included to correct continuum shape differences between the optimal
stellar template and the galaxy spectra due to imperfect flux
calibration and reddening.

Errors on the kinematic measurements were estimated using a set of
Monte Carlo simulations. During each realization, random Gaussian
noise was added to the observed spectrum from each bin based upon the
standard deviation of the pPXF model residuals. The LOSVD was measured
with the penalization turned off in order to produce realistic
uncertainties. From 100 realizations, we determined the 1$\sigma$
errors from the standard deviation of the resulting distributions.

We also tested the robustness of our kinematic measurements by
assuming different continuum models, each with various combinations of
additive and multiplicative polynomials of degree $1 - 3$. We
additionally used pPXF to measure just $V$ and $\sigma$ in each
spatial bin, as well as allowed for the relative contributions of the
stars making up the optimal stellar template to vary between spatial
bins. Regardless of the continuum model used, the number of
Gauss-Hermite moments extracted, or whether a new optimal template was
fit for each bin, we found consistent results for all but a few
spatial bins. The kinematics measured from these few spatial bins were
deemed unreliable and excluded from further analysis.

As a final step, we determined the systematic offsets in the odd
Gauss-Hermite moments directly from the data using the
point-symmetrization routine described in Appendix A of
\cite{vandenBosch_deZeeuw_2010}. The offset for the first odd
Gauss-Hermite moment corresponds to the recession velocity of the
galaxy, and a systematic shift in the second odd Gauss-Hermite moment
may result from slight template mismatch effects. The offsets were
subtracted from the $V$ and $h_3$ values determined with pPXF.

Although the systematic offsets in the odd moments were measured with
the point-symmetrization code, we did not symmetrize the kinematics
before using them as inputs into the orbit-based dynamical
models. Often the observed kinematics are symmetrized
\citep[e.g.,][]{Gebhardt_2003, Cappellari_2006} because the dynamical
models can only produce predicted kinematics that are bi- or
point-symmetric. Thus, symmetrizing the input kinematics reduces the
noise in the observations and aids in visual comparisons between
various models. However, the symmetrization routine assumes that the
galaxy nucleus is exactly centered on one of the IFU lenslets. While
this is true for the reduced data sets from the SAURON IFU
\citep{Bacon_2001}, for which the symmetrization code was originally
designed, the assumption does not hold for the OSIRIS
data. Consequently, our best-fit stellar dynamical model, presented in
\S \ref{sec:results}, was determined using non-symmetrized kinematics,
but in \S \ref{subsec:more_errors} we also test the effect on
$M_\mathrm{BH}$ of symmetrizing the kinematics before running the
orbit-based stellar dynamical models.

\subsection{OSIRIS Kinematics}
\label{subsec:osiris_kin}

\begin{figure}
\begin{center}
\epsscale{1.1}
\plotone{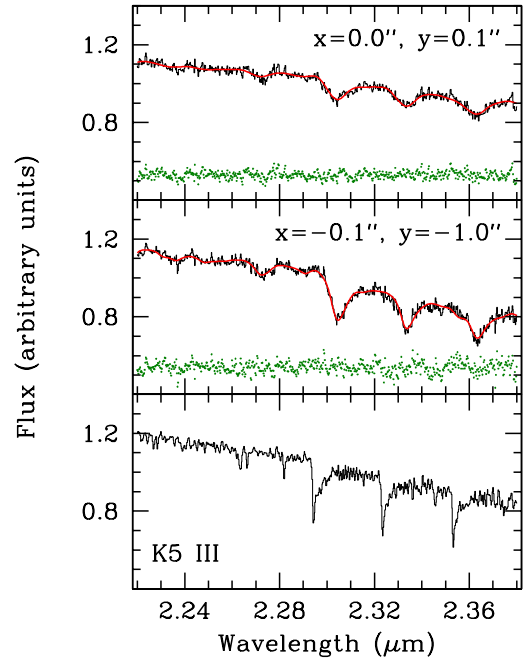}
\caption{The best-fit broadened optimal stellar template (red) is
  compared to the observed NGC 3998 spectrum from a single OSIRIS
  lenslet located at $x = $0\farcs0, $y = $0\farcs01 (top) and a
  spatial bin containing 25 OSIRIS lenslets centered on $x =
  -$0\farcs1, $y = -$1\farcs0 (middle). Green dots show the fit
  residuals and have been shifted by an arbitrary amount. The optimal
  template used to measure the kinematics is dominated by a K5 III
  star, whose spectrum is shown in the bottom
  panel. \label{fig:spec_osiris}}
\end{center}
\end{figure}

We used the pPXF method to measure the kinematics from the OSIRIS data
over a wavelength range of $2.22 - 2.38\ \mu$m. For NGC 3998, this
wavelength region included three very strong CO bandhead absorption
features: $(2-0)^{12}$CO, $(2-1)^{12}$CO, and $(4-2)^{12}$CO. The
kinematics were constrained mainly by these prominent CO bandheads,
although weaker \ion{Ca}{1} and \ion{Mg}{1} absorption lines were also
detected. We note that no emission lines were detected with the OSIRIS
Kbb filter. The stellar kinematics were measured from 90 spatial bins,
10 of which contain just a single lenslet and fall within the central
$\sim$0\farcs15 region. The S/N of the spectra (taken to be the ratio
of median value of the binned spectrum to the standard deviation of
the pPXF model residuals) ranged from $39 - 73$, with a median value
of 58. The optimal template was composed of eight stars, with a K5 III
star contributing a majority of the flux. In Figure
\ref{fig:spec_osiris}, we show example spectra from two spatial bins
and the optimal template broadened by the best-fit LOSVD, as well as
the spectrum from the K5 III star.

The kinematics show that the galaxy is rapidly rotating, with
velocities of $\pm 115$ km s$^{-1}$ over the OSIRIS IFU, and there is
an anti-correlation between $h_3$ and $V$. Although the $h_4$
kinematics are expected to be symmetric about the center, with a peak
or dip at the nucleus, we find $h_4$ to be noisy, without any clear
trends. In contrast, the velocity dispersion shows a very strong peak
at the nucleus, rising from $\sigma \sim$270 km s$^{-1}$ in the outer
regions to $\sigma \sim$560 km s$^{-1}$ at the center. The errors on
the velocity dispersion measurements are quite large at nucleus and
will be discussed further in \S \ref{subsec:more_errors}, but outside
of the innermost lenslets, we find typical errors of 10 km s$^{-1}$,
13 km s$^{-1}$, 0.02, and 0.03, for $V$, $\sigma$, $h_3$, and $h_4$,
respectively.

NGC 3998 contains an AGN, which dilutes the observed stellar features,
and can lead to difficulties in extracting the kinematics near the
nucleus. During the spectral fitting with pPXF, we included an
additive polynomial to account for the AGN dilution. However, we
measured large uncertainties on the kinematics extracted from four of
the innermost lenslets, suggesting that pPXF cannot disentangle the
AGN from the stellar component in a consistent manner. Our final
best-fit model presented in \S \ref{sec:results} was found using the
kinematic measurements and errors from all the OSIRIS bins, however we
further tested the effect on $M_\mathrm{BH}$ when the central
kinematics are excluded from the modeling (see \S
\ref{subsec:more_errors}).

\subsection{STIS Kinematics}
\label{subsec:stis_kin}

\begin{figure}
\begin{center}
\epsscale{1.1}
\plotone{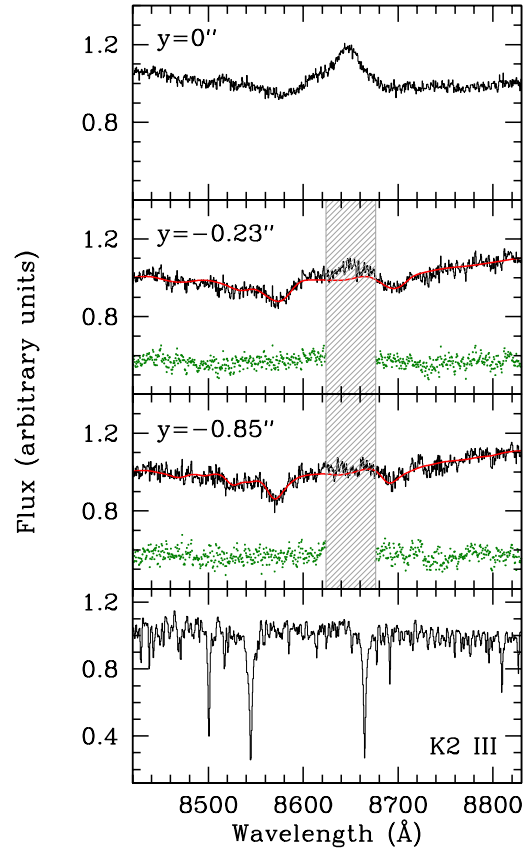}
\caption{STIS spectra extracted from bins located 0\farcs0,
  $-$0\farcs23, and $-$0\farcs85 from the nucleus are shown in the top
  three panels. As can been seen in the top panel, the AGN prevents
  the galaxy's stellar kinematics from being measured near the
  nucleus, and the [\ion{Fe}{2}] emission line is clearly
  detected. Beyond $\sim $0\farcs2 from the nucleus, we were able to
  measure the stellar kinematics, and the middle two panels provide
  examples of the spectral fitting. The best-fit broadened optimal
  template is plotted in red, and the fit residuals are shown by the
  green dots, which have been shifted upward by an arbitrary
  amount. The grey shaded box denotes the wavelength region that was
  excluded during the spectral fitting. The bottom panel shows the
  STIS spectrum of a K2 III star, which contributed most of the flux
  to the optimal stellar template. \label{fig:specfit_stis}}
\end{center}
\end{figure}

We measured the stellar kinematics from the \ion{Ca}{2} triplet lines
using the pPXF routine to fit an optimal stellar template over the
wavelength range $8420 - 8830$ \AA. The optimal template was composed
of two stars: a K2 III star, which contributed most of the flux to the
template, and a K3 III star.  Near the nucleus, there was substantial
AGN contamination, which diluted the the \ion{Ca}{2} triplet lines, as
well as a clear detection of an emission line, which is most likely
the [\ion{Fe}{2}] line at a rest wavelength of 8618 \AA. During the
spectral fitting, we excluded a 50 \AA\ wavelength region surrounding
this emission feature to prevent biases when measuring the
LOSVD. Furthermore, we were unable to measure the stellar kinematics
within $\sim $0\farcs2 of the nucleus due to the AGN. In the end, we
extracted reliable kinematics from eight spatial bins extending from
about 0\farcs2 to 1\arcsec\ from the nucleus. The spectra in the eight
bins had a S/N of $31 - 40$ with a median value of $38$. We present
the nuclear spectrum of NGC 3998, along with examples of the spectral
fitting for two bins from which the stellar kinematics could be
measured in Figure \ref{fig:specfit_stis}.

Like the OSIRIS kinematics, the STIS kinematics show that the galaxy
is rotating quickly with velocities of $\pm 100$ km s$^{-1}$ over the
inner $\pm 1$\arcsec, that there is a rise in velocity dispersion from
$\sigma \sim$ 290 km s$^{-1}$ at 1\arcsec\ to $\sigma \sim$ 370 km
s$^{-1}$ at 0\farcs2, and that $h_3$ is anti-correlated with the
velocity. Again, we are unable to detect any significant trends in
$h_4$. The median errors on the STIS kinematics were 21 km s$^{-1}$,
25 km s$^{-1}$, 0.05, and 0.06 for $V$, $\sigma$, $h_3$, and $h_4$,
and are larger than the uncertainties on the OSIRIS kinematics due to
lower S/N spectra. Given these uncertainties, the STIS data should not
have a substantial impact on the $M_\mathrm{BH}$ determination, and we
found that this was indeed the case when running tests that excluded
the STIS measurements. However, for completeness, the final best-fit
model described in \S \ref{sec:results} was determined using the STIS
data, along with the OSIRIS and LRIS measurements.

\subsection{LRIS Kinematics}
\label{subsec:lris_kin}

\begin{figure}
\begin{center}
\epsscale{1.1}
\plotone{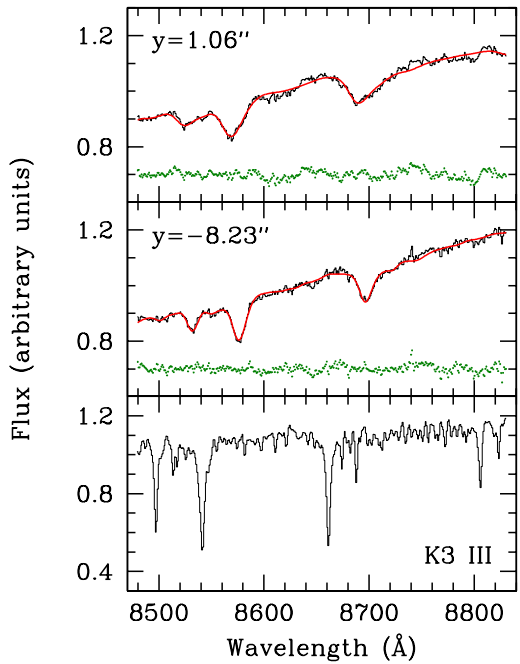}
\caption{Examples of the spectral fitting to the LRIS data. The
  spectra were extracted from bins located 1\farcs06 (top) and
  $-$8\farcs23 (middle) from the nucleus. The optimal stellar
  template, broadened by the best-fit LOSVD, is shown in red, and the
  fit residuals are displayed in green. For comparison, we plot the
  LRIS spectrum of the K3 III star that contributes most of the flux
  to the optimal stellar template. \label{fig:specfit_lris}}
\end{center}
\end{figure}

We used the pPXF method to measure the kinematics from the \ion{Ca}{2}
triplet lines over a wavelength range of 8480 \AA\ $-$ 8830 \AA. The
kinematics in each spatial bin were determined using an optimal
stellar template constructed from the linear combination of a K3 III
and a K0 III star. Ultimately, the kinematics were measured from a
total of 170 spatial bins over the four PAs. The S/N in the bins
ranged from $60 - 107$, with a median value of 82. We present example
fits to the galaxy spectra in addition to the K3 III template in
Figure \ref{fig:specfit_lris}.

We were able to measure the kinematics out to $\sim$15\arcsec\
($\sim$1 kpc) along the major axis and along the intermediate angles
of PA $ = 353^\circ$ and PA $ = 263^\circ$. The kinematics along the
minor axis were measured out to $\sim$8\arcsec. The large-scale
kinematics exhibit features similar to those seen from the
high-resolution OSIRIS maps and STIS data. The stars show rotation
with $V$ between $\pm 200$ km s$^{-1}$, the velocity dispersion
increases steeply toward the center to values of $350 - 400$ km
s$^{-1}$, and $h_3$ is anti-correlated with $V$ across the inner
$\sim$10\arcsec. In addition, we find that $h_4$ is symmetric about
the center with a slight peak to values of $0.05$ at the nucleus. The
median errors for $V$, $\sigma$, $h_3$, and $h_4$ were 8 km s$^{-1}$,
9 km s$^{-1}$, 0.02, and 0.03 for $V$, $\sigma$, $h_3$, and $h_4$.

\begin{figure*}
\begin{center}
\epsscale{0.8}
\plotone{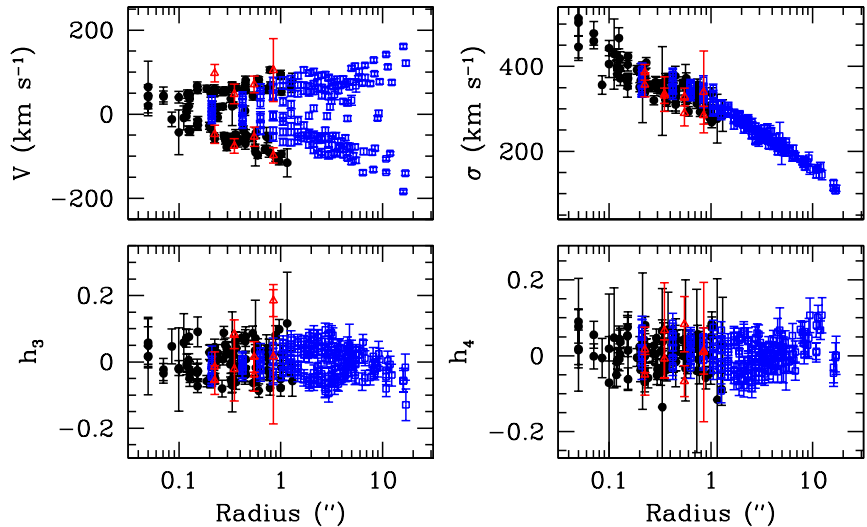}
\caption{A comparison of the stellar kinematics for NGC 3998 measured
  from the OSIRIS (black circles), STIS (red triangles), and LRIS
  (blue squares) data. The radial velocity, $V$, the velocity
  dispersion, $\sigma$, and the higher-order Gauss-Hermite moments,
  $h_3$ and $h_4$, which quantify asymmetric and symmetric deviations
  from a Gaussian, are plotted as a function of projected distance
  from the nucleus. There are two velocity branches in the top left
  plot due to the galaxy's rotation: one side of the galaxy is
  blueshifted relative to the systemic velocity and the other side is
  redshifted. The high angular resolution kinematics from both OSIRIS
  and STIS are in agreement with the large-scale data from
  LRIS. \label{fig:kin_rad}}
\end{center}
\end{figure*}

We find excellent agreement between kinematics measured from the
OSIRIS, STIS, and LRIS data. While template mismatch is a common
source of uncertainty when extracting stellar kinematics, at least it
is encouraging to find consistent measurements from different
instruments over different wavelength regions using different velocity
template stars. In Figure \ref{fig:kin_rad}, we show the four
Gauss-Hermite moments measured from the three different instruments as
a function of projected distance from the nucleus.

\section{PSF Measurements}
\label{sec:psf}

An important input into the stellar dynamical models is the spatial
PSF. However, determining the OSIRIS PSF is especially complicated
because the quality of the AO correction varies with time and numerous
data cubes (obtained over the course of two nights) were mosaiced
together. \cite{Davies_2004} highlight several ways to estimate the
PSF from AO-corrected IFU data, and for this work we used a method
that compares the OSIRIS data cube to a deconvolved \emph{HST} image,
as has been employed by \cite{Krajnovic_2009}. We used the MGE model
generated from the WFPC2 791W image (discussed in \S
\ref{sec:stellarmass}), which already has the \emph{HST} PSF removed,
and convolved it with a model for the OSIRIS PSF. The OSIRIS PSF is
taken to be three concentric circular 2D Gaussians parameterized by
the dispersions and relative weights of the components. The convolved
image is then compared to the OSIRIS data cube, summed along the
wavelength axis, and the parameters of the PSF are varied to determine
the best fit. We found that the three components have dispersions of
0\farcs01, 0\farcs04, and 0\farcs28 with relative weights of 0.19,
0.60, and 0.21. Thus, we were able to achieve a good AO correction and
the PSF core has a FWHM of $\sim$0\farcs1.

To model the STIS PSF, we used Tiny Tim \citep{Krist_Hook_2004} to
reconstruct the PSF for a monochromatic filter passband at 8651
\AA. Fitting two concentric circular 2D Gaussians to the Tiny Tim
image, we found that the PSF can be characterized by a narrow
component with $\sigma = $ 0\farcs03 and a weight of 0.76, plus a
broad component with $\sigma = $ 0\farcs12 and a weight of 0.24. These
dispersion values and weights were used as input into the stellar
dynamical models, which require a PSF to be described with circularly
symmetric Gaussians.

For the LRIS PSF, we examined the spatial distribution of the eight
velocity template stars that were observed on the same night as NGC
3998. We fit a single Gaussian to a spatial cut at 8600 \AA\ through
the geometrically rectified 2D spectra, and found an average
dispersion of $\sigma = $ 0\farcs35.

\section{Orbit-based Stellar Dynamical Modeling}
\label{sec:modeling}

Orbit-based dynamical models using the Schwarzschild superposition
method \citep{Schwarzschild_1979} are the standard means by which to
measure black hole masses from stellar kinematics
\citep[e.g.,][]{Gebhardt_2003, Valluri_2005, Gultekin_2009a,
  Cappellari_2009}. They are additionally useful for studying the
orbital structure of galaxies \citep[e.g.,][]{Verolme_2002,
  Krajnovic_2005, Cappellari_2006, Shen_Gebhardt_2010}, and
constraining dark matter halo properties (e.g., \citealt{Thomas_2005},
2007, \citealt{Weijmans_2009, Forestell_Gebhardt_2010,
  Murphy_2011}). The main strength of this technique is that a
self-consistent distribution function can be constructed from the
observables, without the need for assumptions about the orbital
anisotropy. Here, we calculate three-integral, triaxial, Schwarzschild
models using the implementation of \cite{vandenBosch_2008}, which was
based upon the previous work of \cite{Rix_1997},
\cite{vanderMarel_1998}, \cite{Cretton_1999},
\cite{Verolme_deZeeuw_2002}, and \cite{Cappellari_2006}. We provide a
brief summary of the method here, but refer the reader to
\cite{vandenBosch_2008} for a comprehensive description of model.

In the model, the galaxy potential consists of contributions from the
black hole, the stars, and dark matter. We take the three-dimensional
(3D) stellar mass to be described as the sum of multiple coaxial
Gaussian distributions, which are determined by deprojecting the MGE
model and assuming a constant stellar mass-to-light ratio ($M/L$). The
deprojection is carried out by choosing a direction from which the
galaxy is observed, characterized by the Euler angles $\theta$
(corresponding to the inclination angle), $\phi$, and $\psi$. With a
set of these viewing angles, each component $j$ of the MGE model is
uniquely deprojected into a 3D Gaussian with a shape given by $p_j$,
$q_j$, and $u_j$, which are the ratios between the intermediate and
long axes, the short and long axes, and the length of the longest axis
of the projected Gaussian on the sky to the long axis,
respectively. For details about the deprojection and the relations
between the viewing angles and intrinsic shape parameters see
\cite{vandenBosch_2008}. The Schwarzschild method additionally allows
for dark halo properties to be determined, but our stellar kinematics
extend out to a radius of $\sim $15\arcsec\ ($\sim $1 kpc), and we are
unable to differentiate between various dark halo
models. Consequently, the model results presented in \S
\ref{sec:results} do not include dark matter, and the galaxy potential
is calculated from the contributions of the black hole and the stars
alone. However, we do test the effect on the inferred black hole mass
by incorporating a couple of fixed dark halo models in \S
\ref{subsec:more_errors}.

Given the galaxy potential, a representative orbit library is
generated, and the initial conditions of the orbits are chosen such
that the three integrals of motion are sampled. The orbits are
integrated and projected onto the observable quantities (the location
on the sky and the LOSVD parameters) while accounting for the PSF and
aperture binning. Next, weights for each orbit are found such that
superposition of orbits matches the kinematics in the least-squares
sense while also satisfying contraints imposed by the stellar density
in each aperture and 3D bin. Many dynamical models are calculated with
different values for the parameters of interest, namely
$M_\mathrm{BH}$, $M/L$, and the viewing angles $\theta$, $\phi$, and
$\psi$ (or equivalently the intrinsic shape parameters $p$, $q$, and
$u$). Finally, the best model is taken to be the one that most closely
matches the data in the $\chi^2$ sense.

We fit the Schwarzschild models to the OSIRIS, STIS, and LRIS
kinematics presented in \S \ref{sec:kinematics} and use the stellar
density distribution constructed from the MGE models both with and
without the innermost Gaussian component discussed in \S
\ref{sec:stellarmass}. The MGE model that includes the central
Gaussian component will be discussed in \S \ref{sec:results}, and the
model that assigns all the light from the innermost component to the
AGN will be described in \S \ref{subsec:more_errors}. With four
Gauss-Hermite moments measured in each of the 90 OSIRIS bins, 8 STIS
bins, and 170 LRIS bins, there are 1072 observables. The orbital
libraries were set up to sample 25 equipotential shells at radii
logarithmically spaced from 0\farcs006 to 294\arcsec, with 8 angular
and 8 radial values at each energy to cover the three integrals of
motion. We bundle together $5^3$ orbits with adjacent starting
positions and sum their observables, resulting in a total of 600,000
orbits. Such orbital dithering is commonly used in the construction of
Schwarzschild models to ensure a smooth distribution function, and
further smoothing can be applied after the linear orbital
superposition through the adoption of a regularization term. Our
results presented in \S \ref{sec:results} are based upon models run
without regularization, however we note that including a small amount
of regularization did not change the best-fit black hole mass.

We estimate the model fitting uncertainty on $M_\mathrm{BH}$ ($M/L$)
by marginalizing over $M/L$ ($M_\mathrm{BH}$) and the shape
parameters. Following \cite{Cappellari_2009} and
\cite{Krajnovic_2009}, who advise the use of $3\sigma$ errors for
$M_\mathrm{BH}$ measurements due to the numerical uncertainties
associated with Schwarzschild modeling, we quote the $3\sigma$
uncertainties on $M_\mathrm{BH}$ and $M/L$ for one degree of
freedom. This corresponds to a change of 9.0 in $\chi^2$ from the
minimum value. The statistical uncertainties on the intrinsic shape
parameters are derived differently, using the confidence levels
established by \cite{vandenBosch_vandeVen_2009} and applied during the
studies of M32 and NGC 3379 by \cite{vandenBosch_deZeeuw_2010}. The
confidence interval for the intrinsic shape parameters are set based
upon the expected standard deviation in $\chi^2$, or
$\sqrt{2N_\mathrm{obs}}$, where $N_\mathrm{obs}$ is the number of
observables used to constrain the model (1072 for NGC 3998). While the
$M_\mathrm{BH}$ determination is influenced by a the innermost
kinematical measurements, the intrinsic shape parameters are set by a
much larger number of observables, such that the expected scatter in
$\chi^2$ becomes important and is much larger than the standard
$\Delta \chi^2$ criterion \citep{vandenBosch_vandeVen_2009}.

\section{Results}
\label{sec:results}

It is computationally prohibitive to explore parameter space for the
full range of $M_\mathrm{BH}$, $M/L$, $p$, $q$, and $u$ values
simultaneously, so we initially fixed $M_\mathrm{BH}$ and varied $M/L$
and the shape parameters. We sampled axis ratios of $0.60 \leq p \leq
1.00$, $0.40 \leq q \leq 0.88$, and all possible values of $u$ in
steps of $0.06$, and evaluated 11 $I$-band $M/L$ values between 4.16
and 6.24 $M_\odot/L_\odot$. The minimum values of $p$ and $q$ were
chosen to reflect the smallest values that have been observed in other
galaxies as well as to probe very triaxial shapes, and the upper bound
on $q$ was set by the average flattening of 2D Gaussians in the MGE
model for NGC 3998. With our sampling, we have covered 128 different
galaxy shapes, 5 of which are oblate axisymmetric. This procedure
assumes that the shape parameters do not depend on the black hole
mass, which was shown to be true for the specific cases of M32 and NGC
3379 \citep{vandenBosch_deZeeuw_2010}. We tested whether this also
holds for NGC 3998 by first setting $M_\mathrm{BH} = 2.2 \times 10^8\
M_\odot$ [the \cite{deFrancesco_2006} gas dynamical black hole mass
measurement adjusted for our assumed distance], and then using a
larger mass of $M_\mathrm{BH} = 7.9 \times 10^8\ M_\odot$ [the mass
predicted from the $M_\mathrm{BH} - \sigma_\star$ relationship when a
bulge stellar velocity dispersion of $305$ km s$^{-1}$ is assumed
\citep{Gultekin_2009b}]. We found that the best-fit intrinsic shapes
differ for the two $M_\mathrm{BH}$ values tested: an oblate
axisymmetric shape, with $p = 1.00$ and $q = 0.81$ at an effective
radius ($R_e$) is preferred when $M_\mathrm{BH} = 2.2 \times 10^8\
M_\odot$, and a round, but triaxial shape, with $p = 0.91$ and $q =
0.81$ at $1\ R_e$ is found when $M_\mathrm{BH} = 7.9 \times 10^8\
M_\odot$.

Given that the best-fit intrinsic shape changes with $M_\mathrm{BH}$,
we constructed a model grid by varying $M_\mathrm{BH}$ between $5.0
\times 10^7$ and $5.0 \times 10^9\ M_\odot$ and examining 11 $I$-band
$M/L$ values between 4.2 and 6.2 $M_\odot/L_\odot$, while also
sampling ten shapes. These shapes have the lowest ten $\chi^2$ values
from the two grid runs described above, and they fully encompass the
$3\sigma$ uncertainties of $p$, $q$, and $u$ when $M_\mathrm{BH}$ is
set to $2.2 \times 10^8$ and $7.9 \times 10^8\ M_\odot$. We therefore
should be covering the range of possible shapes for NGC 3998 during
our search for the best-fit $M_\mathrm{BH}$ and $M/L$ parameters.

\begin{figure}
\begin{center}
\epsscale{1.0}
\plotone{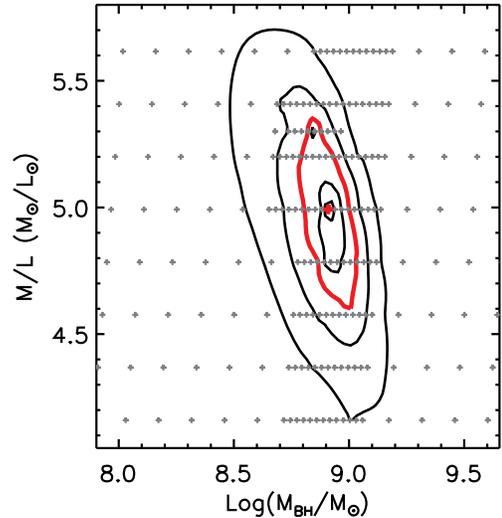}
\caption{Results of the stellar dynamical models after marginalizing
  over the intrinsic shape of the galaxy. At each grey cross, a
  dynamical model was calculated for the specified combination of
  black hole mass and $I$-band $M/L$, and the red cross marks the
  best-fit model. Overplotted are contours of $\Delta \chi^2$, with
  the inner two contours denoting the $1\sigma$ and $2\sigma$
  confidence levels, and the thick red contour signifying the
  $3\sigma$ interval for one degree of freedom. Contours beyond the
  $3\sigma$ confidence level are separated by a factor of
  two. \label{fig:mbhml_contours}}
\end{center}
\end{figure}

\begin{figure*}
\begin{center}
\epsscale{0.82}
\plotone{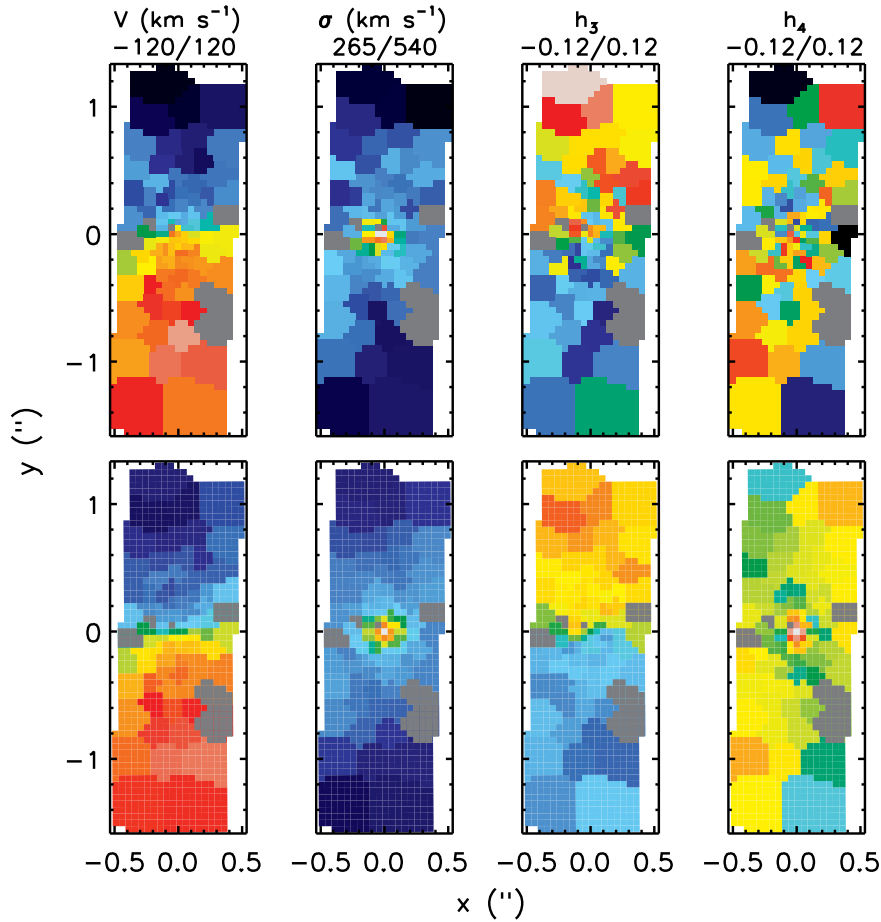}
\caption{The stellar kinematics measured for NGC 3998 from the OSIRIS
  data (top panels) and the predicted values (bottom panels) from the
  best-fit stellar dynamical model with $M_\mathrm{BH} = 8.1 \times
  10^8\ M_\odot$, $I$-band $M/L = 5.0\ M_\odot/L_\odot$, and a shape
  described by $p = 0.96$ and $q = 0.81$ at $1\ R_e$. The velocity map
  shows that the galaxy is rotating rapidly, the velocity dispersion
  map displays a sharp peak within the inner 0\farcs1, and the $h_3$
  map is anti-correlated with the velocity map. The long axis of the
  IFU was aligned with the major axis of the nuclear gas disk at a
  position angle of $308^\circ$, thus the top of the maps correspond
  to the northwest side of the galaxy. The data and model maps are
  plotted on the same scale, with the ranges given by the color bar to
  the right and the minimum and maximum values printed at the top of
  the maps. The kinematics measured from the bins in dark grey were
  deemed unreliable, and were excluded from the subsequent stellar
  dynamical modeling. \label{fig:datamodel_osiris}}
\end{center}
\end{figure*}

\begin{figure}
\begin{center}
\epsscale{1.0}
\plotone{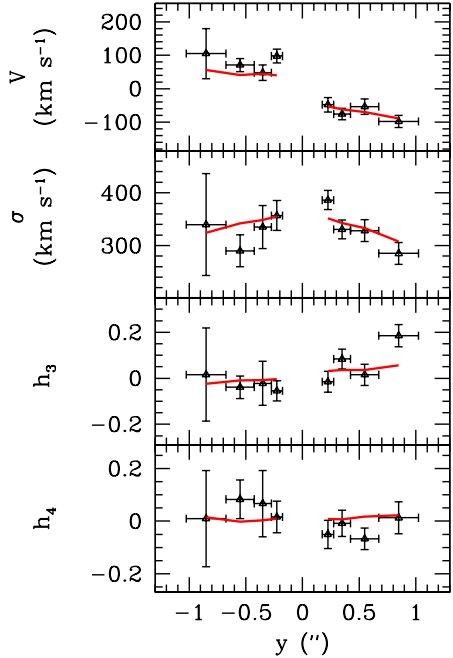}
\caption{The observed STIS kinematics (black) and the predicted
  kinematics (red line) from the best-fit dynamical model. The
  velocity, velocity dispersion, $h_3$, and $h_4$ are plotted as a
  function of position, relative to the nucleus, along the STIS
  slit. The slit was aligned with the galaxy's major axis. We were
  unable to measure the stellar kinematics within $\sim $0\farcs2 of
  the nucleus due to AGN contamination, and model predictions were not
  generated for this region. The kinematics are similar to those
  measured from the OSIRIS data. \label{fig:datamodel_stis}}
\end{center}
\end{figure}

\begin{figure*}
\begin{center}
\epsscale{1.0}
\plotone{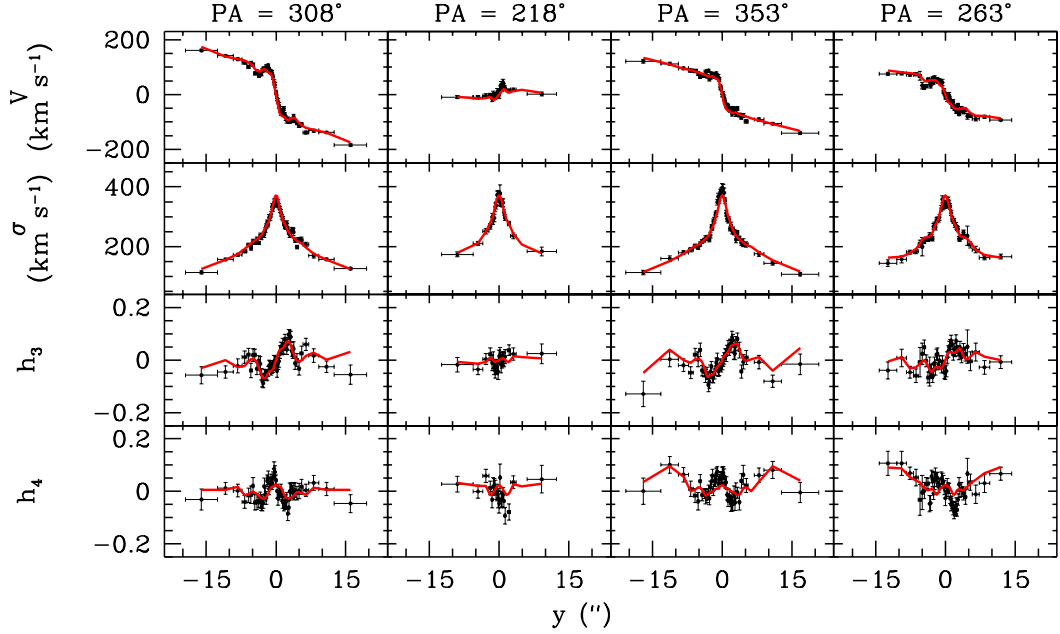}
\caption{The observed LRIS kinematics (black) and predicted values
  from the best-fit dynamical model (red line). The LRIS observations
  were obtained at four position angles: along the major axis of the
  nuclear gas disk (PA $ = 308^\circ$), along the minor axis (PA $ =
  218^\circ$), and at two intermediate angles (PA $ = 353^\circ$ and
  PA $ = 263^\circ$). For each of the slit positions, the velocity,
  velocity dispersion, $h_3$, and $h_4$ are plotted as a function of
  location along the slit relative to the nucleus. These large-scale
  kinematics exhibit similar features to those seen from the high
  angular resolution data, which include rapid rotation, a steep rise
  in the velocity dispersion, and an anti-correlation between $h_3$
  and $V$. Furthermore, $h_4$ is symmetric about the center with a
  slight peak at the nucleus. \label{fig:datamodel_lris}}
\end{center}
\end{figure*}

\begin{figure*}
\begin{center}
\epsscale{0.82}
\plotone{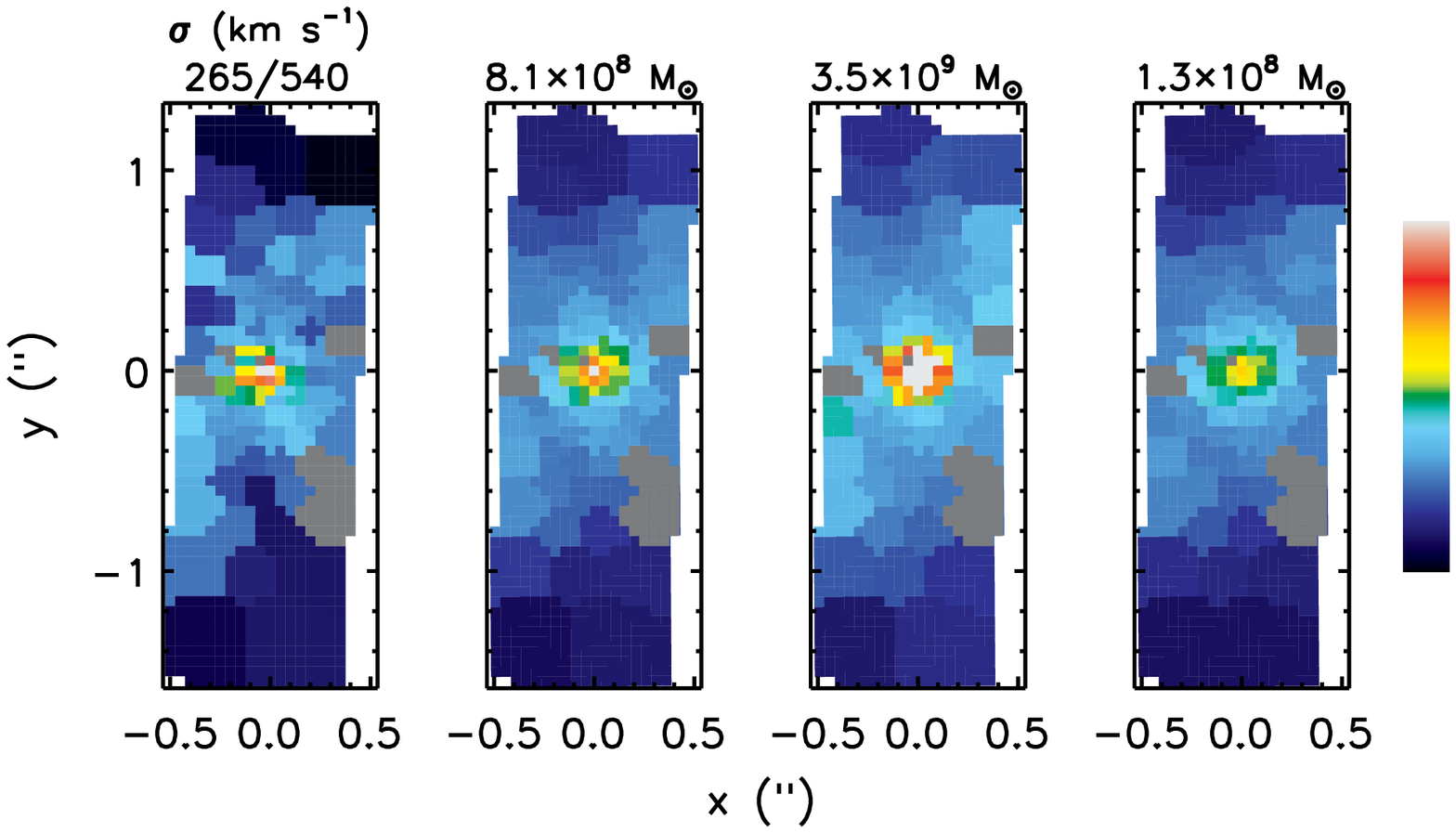}
\caption{Comparison between the observed velocity dispersions over the
  OSIRIS field of view and the predicted ones for various dynamical
  models. The models were constructed with different black hole masses
  (the best-fit $M_\mathrm{BH}$ of $8.1 \times 10^8\ M_\odot$, a
  larger black hole with $M_\mathrm{BH} = 3.6 \times 10^9\ M_\odot$,
  and a smaller black hole of $M_\mathrm{BH} = 1.3 \times 10^8\
  M_\odot$), but have the same $I$-band $M/L$ ($5.0\ M_\odot/L_\odot$)
  and intrinsic shape ($p = 0.96$ and $q = 0.81$ at $1\ R_e$). All
  maps are plotted on the same scale, given by the color bar to the
  right along with the minimum and maximum values listed on top of the
  left-most panel. Observed kinematics measured from the grey bins
  were deemed unreliable, and we do not show the model predictions in
  these bins. \label{fig:osiris_3panel}}
\end{center}
\end{figure*}

The results of our dynamical models are summarized by Figure
\ref{fig:mbhml_contours}, which displays the contours of $\chi^2$ as a
function of $M_\mathrm{BH}$ and $M/L$ after marginalizing over galaxy
shape. The best model has a $\chi^2$ of $1600$, corresponding to a
$\chi^2$ per degree of freedom ($\chi^2_\nu$) of 1.5, with
$M_\mathrm{BH} = (8.1_{-1.9}^{+2.0}) \times 10^8\ M_\odot$, $I$-band
$M/L = 5.0_{-0.4}^{+0.3}\ M_\odot/L_\odot$, and an intrinsic shape
described by $p = 0.96$ and $q = 0.81$ at $1\ R_e$. The
$M_\mathrm{BH}$ and $M/L$ errors represent the $3\sigma$ statistical
uncertainties, and the errors for the shape parameters will be
discussed below. The kinematics predicted from such a model are
compared to the observed OSIRIS, STIS, and LRIS data in Figures
\ref{fig:datamodel_osiris}, \ref{fig:datamodel_stis}, and
\ref{fig:datamodel_lris}, respectively. We additionally compare the
OSIRIS velocity dispersions to those predicted from models with the
best-fit mass of $M_\mathrm{BH} = 8.1 \times 10^8\ M_\odot$, a smaller
black hole with $M_\mathrm{BH} = 1.3 \times 10^8\ M_\odot$, and a
larger black hole with $M_\mathrm{BH} = 3.6 \times 10^9\ M_\odot$ in
Figure \ref{fig:osiris_3panel}. This figure demonstrates that our
best-fit black hole mass is a reasonable one, and clear differences
between the predicted velocity dispersions for the models with a
smaller and larger black hole and the observed kinematics can be seen
by eye. The more massive black hole produces velocity dispersions that
are much too large at the center, while the smaller black hole is
unable to match the sharp rise in the observed nuclear velocity
dispersion.

\begin{figure}
\begin{center}
\epsscale{1.0}
\plotone{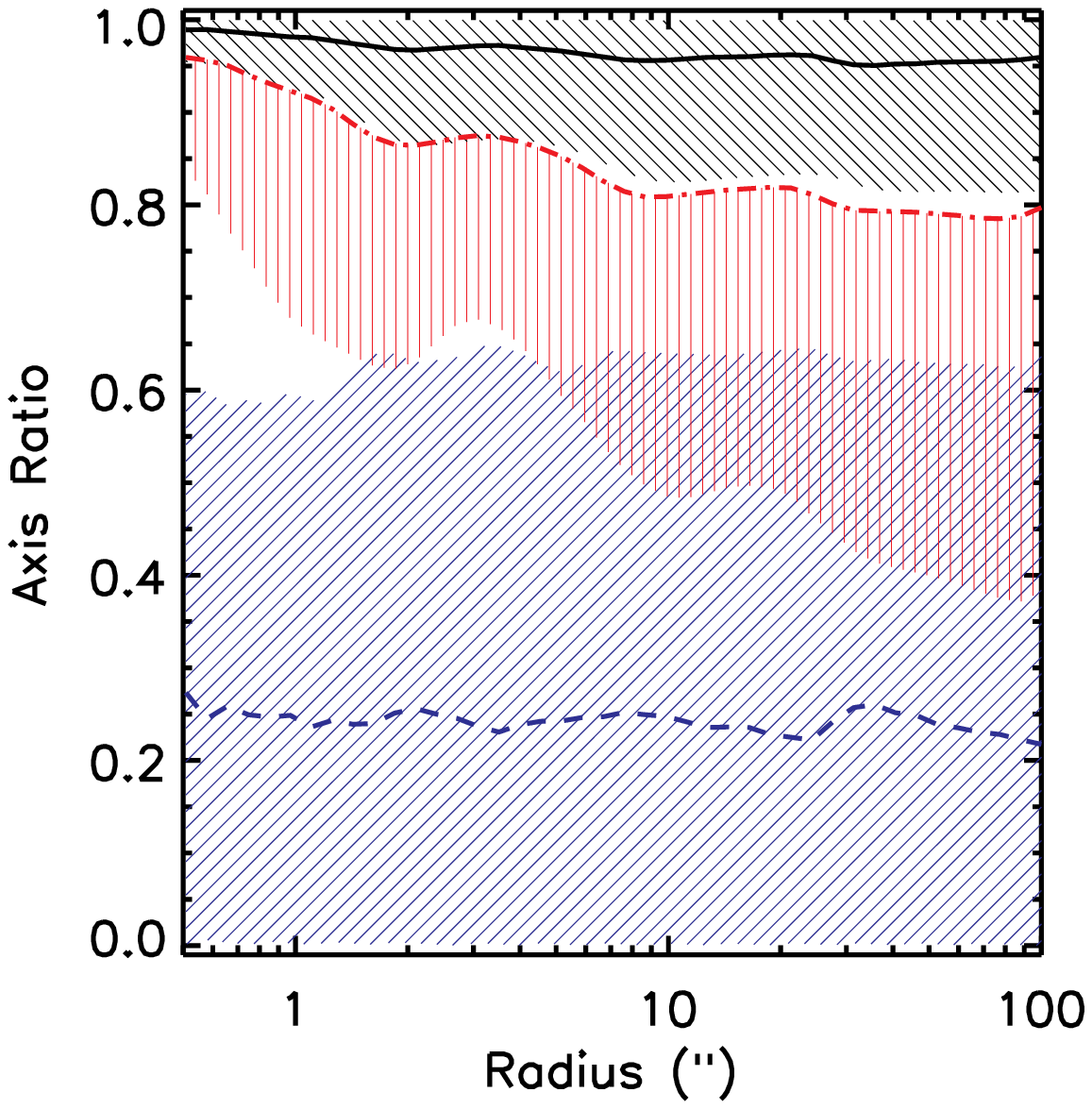}
\caption{The intrinsic shape of NGC 3998. The ratio, $p$, between the
  intermediate and long axes (solid black line), the ratio, $q$,
  between the short and long axes (dash-dotted red line), and the
  triaxiality parameter $T$ (dashed blue line) are plotted as a
  function of distance from the nucleus. The $3\sigma$ uncertainties
  are given by the range covered by the black backward, red vertical,
  and blue forward slashes for $p$, $q$, and $T$, respectively. The
  kinematical measurements from the large-scale LRIS data extend out
  to a radius of $\sim$15\arcsec. \label{fig:axisratio_radius}}
\end{center}
\end{figure}

We calculated another model grid holding $M_\mathrm{BH}$ fixed at
$M_\mathrm{BH} = 8.1 \times 10^8\ M_\odot$ and allowing $M/L$ and the
shape parameters to vary. This grid allowed us to derive the
uncertainties on the shape parameters, having previously determined
the best-fit black hole mass. We again sampled 128 galaxy shapes with
axis ratios of $0.60 \leq p \leq 1.00$, $0.40 \leq q \leq 0.88$, and
all possible values of $u$, and examined 11 $I$-band $M/L$ values
between 4.2 and 6.2 $M_\odot/L_\odot$. We found that the galaxy shape
can be described with the ratios $p = 0.96_{-0.13}^{+0.04}$ and $q =
0.81_{-0.33}^{+0.00}$ ($3\sigma$ uncertainties) at $1\ R_e$. In Figure
\ref{fig:axisratio_radius}, we plot the best-fit axis ratios $p$ and
$q$ and their uncertainties at all radii, extending out to
100\arcsec. We further show the radial variation of the triaxiality
parameter, $T = (1-p^2)/(1-q^2)$, on the same plot. An upper error bar
of zero is measured for $q$ because this ratio is limited by the
observed average flattening of the 2D Gaussians from the MGE
model. Thus, the best-fit intrinsic shape is as round as the
\emph{HST}/CFHT images allow, and is consistent with an oblate
axisymmetric spheroid. Furthermore, we were unable to place strong
constrains on the viewing angles, finding that the inclination ranges
from $\theta = 38^\circ$ to $\theta = 90^\circ$ (edge-on), $\phi =
-90_{-0}^{+82}$, and $\psi = 90_{-4}^{+0}$ ($3\sigma$
uncertainties). These results are consistent with previous stellar
dynamical studies of other early-type galaxies, in particular the work
of \cite{Krajnovic_2005} and \cite{vandenBosch_vandeVen_2009} who find
that the viewing angles are highly degenerate.

\subsection{Additional Sources of Uncertainty}
\label{subsec:more_errors}

The $3\sigma$ errors on $M_\mathrm{BH}$ presented above represent the
formal model fitting uncertainty, and account for the random noise
within the stellar dynamical models. We further explored other sources
of uncertainty that are not included in the statistical errors but
that could have an effect the $M_\mathrm{BH}$ determination. We
summarize the results of these model grids below.

\emph{Stellar Density Distribution}: Care must be taken to separate
the AGN light from the stellar contribution when constructing the
luminous mass model. Often, this is accomplished by removing the
innermost Gaussian component from the MGE model, however NGC 3998
exhibits a steep surface brightness profile and some starlight may
still be contained within the central component. In the analysis
presented above, we assumed that all of the light from the central MGE
Gaussian was due to the stars. Here, we consider the other extreme,
where all the light from the inner MGE component comes from the AGN.

After removing the central Gaussian component from the MGE model, we
followed the same procedure outlined in \S \ref{sec:results}. We first
calculated stellar dynamical models fixing $M_\mathrm{BH}$ to $2.2
\times 10^8\ M_\odot$ and $7.9 \times 10^8\ M_\odot$ in order to
determine the most probable galaxy shapes. Then, we ran model grids
sampling over the ten best galaxy shapes, while varying
$M_\mathrm{BH}$ between $5.0 \times 10^7$ and $5.0 \times 10^9\
M_\odot$ and $M/L$ between 4.2 and 6.2 $M_\odot/L_\odot$. We found a
very round, but oblate, intrinsic shape for the galaxy, with
$M_\mathrm{BH} = (10.1_{-3.5}^{+0.7}) \times 10^8\ M_\odot$ and an
$I$-band $M/L = 4.8_{-0.1}^{+0.6}\ M_\odot/L_\odot$ ($3\sigma$
uncertainties). This best-fit stellar dynamical model is a worse
description of the observations ($\chi^2=1619$) compared to the best
model presented previously in \S \ref{sec:results}
($\chi^2=1600$). Moreover, when excluding the central MGE component,
the dynamical models for every combination of $M_\mathrm{BH}$ and
$M/L$ have a higher $\chi^2$ than the stellar dynamical models
constructed from a mass model with all 12 Gaussian components. For
this reason, and because the black hole masses determined using the
two MGE models are fully consistent within the statistical
uncertainties, covering nearly identical $M_\mathrm{BH}$ ranges, we
view the best-fit model from \S \ref{sec:results} as our final result.

\emph{Dark Halo}: Work by \cite{Gebhardt_Thomas_2009} has raised
concerns that stellar dynamical black hole mass measurements may be
underestimated if both the black hole and the dark matter halo are not
simultaneously modeled. The reasoning is that without a dark halo, the
$M/L$ may be overestimated in order to compensate for the missing mass
at large radii. The $M/L$ is assumed to be constant in the models, and
thus a smaller black hole would be needed to match the observed
nuclear kinematics. By including the contribution of dark matter in
their re-examination of M87, \cite{Gebhardt_Thomas_2009} found that
the black hole mass increased by about a factor of two. Likewise,
\cite{McConnell_2011a} noted a strong dependence of the black hole
mass on the dark halo when studying NGC 6086. Both objects exhibit
shallow stellar luminosity profiles and have massive, concentrated
dark matter halos, limiting the radial range over which the stars
dominate the gravitational potential and the assumption of a constant
$M/L$ holds. Moreover, these studies utilized data in which the
quality and/or spatial resolution was unable to limit the degeneracy
between the black hole and stellar mass-to-light ratio in the central
regions. In contrast, black hole mass measurements based upon high S/N
observations well within the region influenced by the black hole
appear to be less sensitive to the inclusion of dark halos during the
modeling \citep{Shen_Gebhardt_2010, Schulze_Gebhardt_2011,
  Gebhardt_2011}.

We are unable to directly fit for the parameters of a dark halo model
because our kinematics do not extend out to large enough
radii. Instead, we selected a fixed model for the dark halo. Two
commonly used dark halo models are the Navarro-Frenk-White (NFW)
profile \citep{NFW} and a distribution based on a cored logarithmic
potential \citep{Binney_Tremaine_1987, Thomas_2005}. Previous work has
found that both parameterizations give consistent results
\citep{Thomas_2007, Gebhardt_Thomas_2009, McConnell_2011a}, and here
we adopt the profile from a logarithmic potential given by

\begin{equation}
\label{eq:logdm}
\rho_\mathrm{DM}(r) = \frac{V_c^2}{4 \pi G} \frac{3r_c^2 + r^2}{(r_c^2
  + r^2)^2} .
\end{equation}

\noindent The parameters $V_c$ and $r_c$ are the asymptotically
constant circular velocity and the core radius, within which the dark
matter density is approximately constant. We fixed $V_c$ and $r_c$ to
values of $407$ km s$^{-1}$ and $10.7$ kpc, respectively, which were
chosen for NGC 3998 using the galaxy $B$-band luminosity of $1.6
\times 10^{10}\ L_\odot$ reported by the Hyperleda database
\citep{Paturel_2003} and the empirical relations by
\cite{Thomas_2009}. The relationships in \cite{Thomas_2009} provide a
way to select physically motivated parameters for the dark matter
halo, and was similarly used by \cite{Schulze_Gebhardt_2011} to test
the effects of dark matter on the black hole mass.

When including the dark halo in our stellar dynamical models, we did
not observe a significant change in the black hole mass, measuring
$M_\mathrm{BH} = (7.1_{-0.9}^{+3.7}) \times 10^8\ M_\odot$ and an
$I$-band $M/L = 5.0_{-0.4}^{+0.2}\ M_\odot/L_\odot$ ($3\sigma$
errors). The decrease in the best-fit black hole mass is in the
opposite direction as anticipated, but is well within the statistical
uncertainties and is likely due to noise in the dynamical models. A
similar result was seen when we incorporated a dark halo that was
twice as massive within the radial range covered by the large-scale
LRIS kinematics. Therefore, neglecting the dark matter contribution
during the dynamical modeling presented in \S \ref{sec:results} did
not lead to a biased $M_\mathrm{BH}$ determination. We attribute this
result to the high quality OSIRIS data, which probes regions well
within the influence of the black hole and reduces the degeneracy
between $M_\mathrm{BH}$ and $M/L$, as well as to the galaxy's strong
rotation, which also makes the measurement less sensitive to the dark
halo.

\emph{Central OSIRIS Kinematics}: Contributions from a non-stellar
nucleus dilute absorption lines and make the measurement of reliable
stellar kinematics challenging. In some instances, the AGN
contribution is so strong that the stellar features are no longer
visible in the nuclear spectra \citep{Cappellari_2009}. This is not
the case for NGC 3998, however we do measure large errors for the
kinematics extracted from four of innermost OSIRIS lenslets. As an
example, at these locations we found velocity dispersions between
$500$ and $560$ km s$^{-1}$ with error bars of $42 - 146$ km
s$^{-1}$. Beyond these four lenslets, but still within $\sim$0\farcs1
of the nucleus, we continued to measure large velocity dispersions of
$\sigma \sim 450$ km s$^{-1}$ but with more reasonable errors of
$\sim$20 km s$^{-1}$. The large uncertainties from the inner OSIRIS
bins indicate that the pPXF routine could not model the AGN
contribution in a consistent fashion within 0\farcs05 from the
nucleus, which in turn may have biased these kinematical
measurements. Therefore, we excluded the measurements from the central
five OSIRIS bins when fitting the orbit-based stellar dynamical
models. Doing so led to a best-fit model with $M_\mathrm{BH} = (8.1
\pm2.0) \times 10^8\ M_\odot$ and $I$-band $M/L = 5.0_{-0.4}^{+0.2}\
M_\odot/L_\odot$. The values and $3\sigma$ uncertainties for
$M_\mathrm{BH}$ and $M/L$ are the same as found previously in \S
\ref{sec:results}, when all of the OSIRIS kinematics were fit during
the modeling. Even if the kinematics measured from four of the
innermost OSIRIS lenslets are biased due to AGN contamination, there
appears to be no subsequent effect on the black hole mass. This result
is not entirely surprising because the stellar dynamical models take
into account the errors on the input kinematics, and we have a number
of other reliable measurements of the LOSVD within the black hole
sphere of influence.

\emph{OSIRIS PSF}: As discussed in \S \ref{sec:psf}, the OSIRIS PSF
was determined by comparing the deconvolved \emph{HST} WFPC2 F791W
image to the OSIRIS $K$-band data cube summed along the wavelength
axis. This approach assumes that the comparison image has a higher
resolution than the collapsed data cube and that there is no strong
color gradient. Given the difficulties associated with measuring the
PSF, we therefore additionally considered two extreme
parameterizations of the OSIRIS PSF to quantify changes in
$M_\mathrm{BH}$. Both PSFs were composed of two concentric, circular,
2D Gaussians, and the dispersion of the second component was set at
0\farcs28. We chose 0\farcs28 because this is the dispersion of the
broad component in the OSIRIS PSF model presented in \S
\ref{sec:psf}. We then constructed a ``poor'' PSF, such that the
narrow component had a dispersion of 0\farcs084 and contributed 15\%
of the flux to the total PSF, and a ``good'' PSF, whose narrow
component had a dispersion of 0\farcs025 (roughly the $K$-band
diffraction limit for a 10m telescope) and contributed 95\% of the
flux to the total PSF. The stellar dynamical models based upon the
``bad'' PSF yielded a best-fit model with $M_\mathrm{BH} =
(8.7_{-1.6}^{+3.9}) \times 10^8\ M_\odot$ and $I$-band $M/L =
5.0_{-0.4}^{+0.2}\ M_\odot/L_\odot$, and the models using the ``good''
PSF produced best-fit parameters of $M_\mathrm{BH} =
(8.1_{-1.4}^{+1.1}) \times 10^8\ M_\odot$ and $I$-band $M/L =
5.0\pm0.2\ M_\odot/L_\odot$. Thus, even for two extreme PSF models,
the black hole mass remains within the $3\sigma$ statistical
uncertainties presented in \S\ref{sec:results}.

\emph{Symmetrizing the Kinematics}: We removed the systematic offsets
in the odd Gauss-Hermite moments (e.g., the galaxy's systemic
velocity) with the point-symmetrization code of
\cite{vandenBosch_deZeeuw_2010}. In addition to this step, the common
practice is to make further adjustments to the kinematical
measurements and their errors so that the velocity fields become bi-
or point-symmetric, thereby reducing the observational noise. However,
the symmetrization process assumes that the galaxy nucleus is centered
on one of the lenslets, which is not necessarily true for the OSIRIS
data. Therefore, in \S \ref{sec:results}, we calculated stellar
dynamical models without first requiring the observed velocity fileds
to be symmetric. Instead, if we do symmetrize the input kinematics, we
find a best-fit model with $M_\mathrm{BH} = (9.5_{-3.2}^{+2.1}) \times
10^8\ M_\odot$ and an $I$-band $M/L = 4.8_{-0.2}^{+0.4}\
M_\odot/L_\odot$. As expected, the best model is a much better
description of the (symmetrized) observed kinematics, with $\chi^2 =
459$ and $\chi^2_\nu = 0.43$, and larger statistical $3\sigma$
uncertainties on $M_\mathrm{BH}$ are found. Although the best-fit
black hole mass increased by 17\% from the mass derived previously, it
remains within the model fitting uncertainties determined in \S
\ref{sec:results}, and we continue to prefer the dynamical model
constrained with the non-symmetrized kinematics.

\section{Discussion}
\label{sec:discussion}

Through orbit-based stellar dynamical modeling we measured a mass of
$M_\mathrm{BH} = (8.1_{-1.9}^{+2.0}) \times 10^8\ M_\odot$ for the
black hole in NGC 3998. We report $3\sigma$ statistical errors on the
black hole mass, but ran additional model grids to assess the
robustness of our mass measurement. We found no significant changes
(outside the random noise of the models) to the black hole mass due to
these sources of uncertainty, and so we do not make any additional
adjustments to the black hole mass or its error bars. With this black
hole mass, and a bulge stellar velocity dispersion of 272 km s$^{-1}$
(see \S \ref{subsec:mbh_correlations} below), the black hole sphere of
influence is $r_\mathrm{sphere} = $ 0\farcs7, which is well resolved
by the observations. We now examine the orbital structure of galaxy,
compare the stellar dynamical black hole mass measurement to the
existing gas dynamical mass, and place NGC 3998 on the $M_\mathrm{BH}$
$-$ host galaxy relationships.

\subsection{Orbital Structure}
\label{subsec:orbital_structure}

\begin{figure}
\begin{center}
\epsscale{1.1}
\plotone{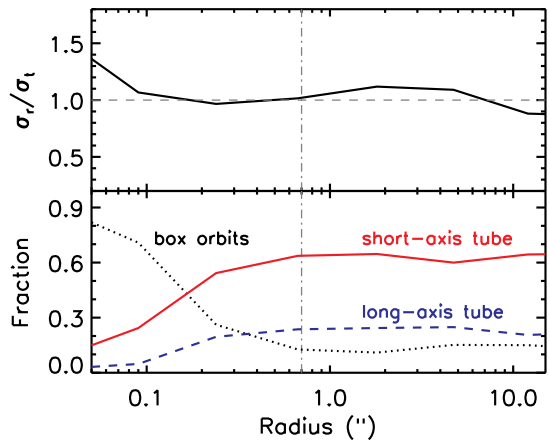}
\caption{Orbital structure of NGC 3998 from our best fitting stellar
  dynamical model. The anisotropy, $\sigma_r/\sigma_t$, (top) and the
  orbit type (bottom) are plotted a s function of radius, covering the
  extent of the kinematic measurements. The vertical grey dot-dashed
  line designates the black hole sphere of influence. NGC 3998 is
  mostly isotropic (indicated by the horizontal dashed grey line in
  the top panel) but shows a radial bias within $\sim$0\farcs1. Near
  the nucleus, box orbits (black dotted line) dominate the galaxy,
  while short-axis tube orbits (red solid line) become important at
  larger radii. Long-axis tube orbits are shown by the blue dashed
  line. \label{fig:orbstructure}}
\end{center}
\end{figure}

We examined the internal orbital structure of NGC 3998 using the
orbital weights found with the Schwarzschild superposition method for
our best fitting model (presented in \S \ref{sec:results}). Defining
the tangential velocity dispersion as $\sigma_t^2 = (\sigma_\phi^2 +
\sigma_\theta^2)/2$, with $(r, \theta, \phi)$ being the usual
spherical coordinates, we plot the ratio $\sigma_r/\sigma_t$ as a
function of radius in the top panel of Figure
\ref{fig:orbstructure}. Near the nucleus, the velocity dispersion
becomes radially anisotropic reaching values of $\sigma_r/\sigma_t
\sim 1.5$, while the velocity dispersion is isotropic, deviating by at
most $\sim$10\% from $\sigma_r/\sigma_t = 1$, at radii beyond
$\sim$0\farcs1. The radial anisotropy near the galaxy's center can be
attributed to the large fraction of box orbits: we found that the box
orbits contribute $60 - 80$\% to the stellar mass within the inner
$\sim$0\farcs1 (see bottom panel of Figure
\ref{fig:orbstructure}). Similarly large contributions from box orbits
have been seen when applying triaxial Schwarzschild models to other
objects as well \citep{Weijmans_2009,
  vandenBosch_deZeeuw_2010}. Outside of $r_\mathrm{sphere}$, our
best-fit model is made up of $\sim$65\% short-axis tube orbits and
$\sim$20\% long-axis tube orbits.

\begin{figure}
\begin{center}
\epsscale{1.1}
\plotone{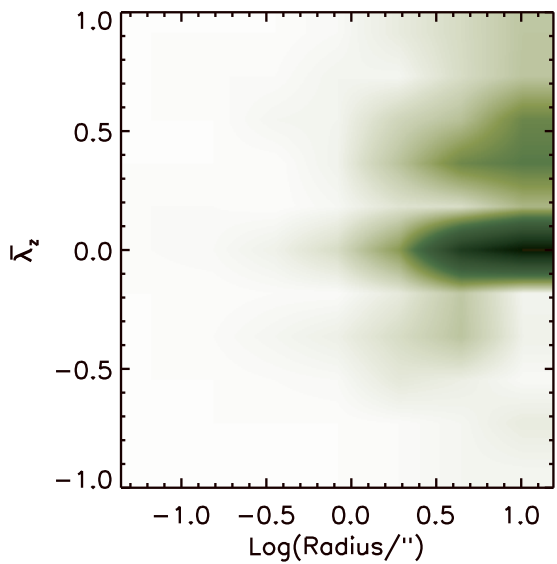}
\caption{Distribution of mass along the orbits from our best fitting
  dynamical model as a function of average radius and spin. The radial
  range extends over the region covered by our kinematic
  measurements. A non-rotating bulge ($-0.2 < \bar \lambda_z < 0.2$)
  and rotating disk ($\bar \lambda_z \geq 0.2$) component are clearly
  present. \label{fig:dynamical_bulgedisk}}
\end{center}
\end{figure}

NGC 3998 is classified as an S0 galaxy, and indeed, our best fitting
dynamical model shows evidence for both a bulge and a disk
component. In Figure \ref{fig:dynamical_bulgedisk}, we show the mass
distribution along orbits as a function of average radius and spin,
$\bar \lambda_z$. The spin is defined as $\bar \lambda_z = \bar J_z
\times (\bar r / \bar \sigma)$, where $\bar J_z$ is the average
angular momentum along the $z$-direction, $\bar r$ is the average
radius, and $\bar \sigma$ is the average second moment of the
orbit. The figure shows the clear presence of a non-rotating bulge
component ($-0.2 < \bar \lambda_z < 0.2$) and a rotating disk
component ($\bar \lambda_z \geq 0.2$), making up 52\% and 39\% of the
mass within the radial range covered by our kinematic
measurements. The remaining 9\% of the mass comes from orbits with
$\bar \lambda_z \leq -0.2$ that fall outside the radial range covered
by the OSIRIS kinematics. Thus, the near isotropy at radii larger than
$\sim$0\farcs1 seen in Figure \ref{fig:orbstructure} is due to the
presence of a bulge component, while the disk causes the strong
rotation that is seen in the OSIRIS, STIS, and LRIS kinematics in
Figures \ref{fig:datamodel_osiris}, \ref{fig:datamodel_stis}, and
\ref{fig:datamodel_lris}.

\subsection{Comparison to the Gas Dynamical Measurement}
\label{subsec:gas_compare}

By modeling the \emph{HST} STIS gas kinematics as a thin, circularly
rotating disk, \cite{deFrancesco_2006} derived a black hole mass of
$M_\mathrm{BH} = (2.7_{-2.0}^{+2.4}) \times 10^8\ M_\odot$ ($2\sigma$
uncertainties). Scaling the black hole mass for the distance adopted
here, their mass measurement becomes $M_\mathrm{BH} =
(2.2_{-1.6}^{+1.9}) \times 10^8\ M_\odot$. The stellar dynamical black
hole mass is therefore significantly larger, by nearly a factor of
four, compared the gas dynamical measurement. We cannot resolve the
discrepancy between the stellar and gas dynamical mass measurements
for the black hole in NGC 3998, but discuss a few of the uncertainties
associated with both methods below.

The gas kinematics measured from the STIS data clearly show organized
rotation. However, the line widths predicted from a rotating,
dynamically cold, thin-disk model do not fully match the observed
velocity dispersions. \cite{deFrancesco_2006} found that the model
underestimates the observed rise in the nuclear velocity dispersion by
$\sim$20\%, although larger inconsistencies were seen in the two slit
positions immediately adjacent to the central slit. Without
re-analyzing the STIS data for NGC 3998, it is difficult to determine
whether the addition of an intrinsic velocity dispersion to the models
would greatly improve the agreement between the predicted and observed
line widths, and by how much the gas dynamical black hole mass
measurement would then increase if the excess velocity dispersion was
considered dynamically important. Previous gas dynamical studies that
have assigned a dynamical origin to the internal velocity dispersion
have found varying degrees to which the black hole mass is
affected. For example, using an asymmetric drift correction to
estimate the circular velocity from the observed mean rotational
speed, \cite{Barth_2001} found a 12\% increase in the best-fit black
hole mass for NGC 3245 while \cite{Walsh_2010} found that the mass of
the black hole in M84 doubles. However, we note that the differences
between the observed line widths and the predictions from a
dynamically cold, rotating disk model for NGC 3245 and M84 were larger
than those observed for NGC 3998. Admittedly, the factor of four
increase needed to make the best-fit gas dynamical mass to match the
best-fit stellar dynamical measurement is quite large, and the
discrepancy probably cannot be entirely resolved by assigning a
dynamical origin to an intrinsic velocity dispersion.

Another source of uncertainty in the gas dynamical black holes mass
comes from the inclination of the emission-line
disk. \cite{deFrancesco_2006} found that all models with inclinations
below $70^\circ$ produced equally acceptable fits to the data at the
$2\sigma$ level, and that the inferred black hole masses differed by
about an order of magnitude for inclination angles $i = 10^\circ$ and
$70^\circ$. A similarly large difference in the NGC 3998 black hole
mass for extreme inclination angles was reported by
\cite{Beifiori_2009} , who estimated masses of $M_\mathrm{BH} = 3.4
\times 10^8\ M_\odot$ and $4.9 \times 10^7\ M_\odot$ (adjusted to our
distance) for inclinations of $i = 33^\circ$ and
$88^\circ$. \cite{deFrancesco_2006} ruled out very face-on
orientations, $i < 27^\circ$, because such angles produced $V$-band
mass-to-light ratios that are too high compared the predictions of a
Salpeter initial mass function (IMF) for an old stellar
population. They ultimately reported results using a disk inclination
of $30^\circ$, and further combined their statistical uncertainties on
$M_\mathrm{BH}$ with the uncertainties resulting from the allowed
range in inclination angles between $27^\circ$ and $70^\circ$ to
derive the $2\sigma$ error bars on $M_\mathrm{BH}$. Interestingly, the
results of \cite{Cappellari_2012} indicate that the most massive
galaxies have $M/L$ values that are higher than the predictions of a
Salpeter IMF for an old stellar population, thus lower inclinations of
the gas disk in NGC 3998 may be acceptable, which would then lead to
an increase in the gas dynamical mass measurement.

When constructing the stellar dynamical models presented here, we did
not assume axisymmetry or a specific viewing orientation. However, it
is not feasible to explore all possible values for the five parameters
simultaneously, and we employed a method that begins by holding fixed
$M_\mathrm{BH}$ while varying the other parameters to find the
best-fit intrinsic galaxy shape before searching for the best-fit
$M_\mathrm{BH}$. Such an approach does not guarantee that a global
minimum will be found, and we attempted to alleviate this problem by
sampling over ten galaxy shapes (instead of just the best-fit shape)
while varying $M_\mathrm{BH}$ and $M/L$. The ten shapes were taken
from the ten best models when $M_\mathrm{BH}$ was held at two
different values, and the shapes cover the $3\sigma$ ranges of $p$,
$q$, and $u$ for the two black hole masses. Thus, we should have
covered the possible galaxy shapes while searching for the best
fitting $M_\mathrm{BH}$ and $M/L$ parameters.

One assumption that is made when constructing the stellar dynamical
models is that the $M/L$ does not vary with radius. NGC 3998 has both
bulge and disk components, and could also contain a nuclear star
cluster \citep{GonzlezDelgado_2008, Neumayer_2012}. If each of these
components has a very different stellar population, then a single
$M/L$ is not a realistic description of the system. Recent work
\citep[e.g.,][]{vanDokkum_Conroy_2010} has advocated for a
bottom-heavy IMF in massive early-type galaxies. Thus, if the bulge
component of NGC 3998 is indeed rich in dwarf stars, the larger $M/L$
for the bulge region would lead to a smaller stellar dynamical black
hole mass measurement. However, allowing for a radial variation in
$M/L$, in addition to sampling the $M_\mathrm{BH}$ and three shape
parameters, is too computationally expensive to be attempted here. In
contrast, the gas dynamical black hole mass measurement, which relies
on the gas kinematics at small radii where the stellar potential is
dominated by the bulge, is insensitive to $M/L$ variations. While our
assumption of a constant $M/L$, may effect the derivation of the
stellar dynamical black hole mass, the other systematic effects often
associated with the stellar dynamical method appear to be minimal. For
example, we found the black hole mass was insensitive to the inclusion
of a of dark halo in the model, as well as to the exact OSIRIS PSF
that was adopted. Moreover, we found good agreement between the
kinematics measured from different instruments using different
template libraries, suggesting that template mismatch is not a
dominate source of uncertainty.

\subsection{The Black Hole Scaling Relations}
\label{subsec:mbh_correlations}

When placing NGC 3998 on the $M_\mathrm{BH} - \sigma_\star$ relation,
\cite{Gultekin_2009b} adopted a bulge stellar velocity dispersion of
$305$ km s$^{-1}$. The value was taken from the Hyperleda database,
and was based upon the dispersion values found in the literature at
that time, however we are in a position to directly measure the
effective stellar velocity dispersion ($\sigma_\mathrm{e}$) of the
bulge component from our data and models. Recent studies of very
massive black holes have elected to exclude data within
$r_\mathrm{sphere}$ when measuring the effective velocity dispersion
\citep{Gebhardt_2011, Jardel_2011, McConnell_2011b}, and here we
follow the same approach. We weight the $\sigma$ and $V$ measurements
obtained from the LRIS major axis slit with the surface brightness
profile set by our MGE model, following \cite{Gultekin_2009b}, but
integrate from $r_\mathrm{sphere}$ to the bulge effective radius. The
bulge effective radius is rather uncertain for NGC 3998, with
literature values ranging between 4\farcs7 < $R_e$ < 18\farcs3
\citep{Sani_2011, Baggett_1998}, which in turn produces some
uncertainty in the measurement of $\sigma_\mathrm{e}$. Taking an
average of the $R_e $ measurements determined from recent optical
images \citep{Fisher_1996, SanchezPortal_2004}, we find that $R_e =
$10\farcs7 and $\sigma_\mathrm{e} = 272$ km s$^{-1}$ for NGC 3998. For
comparison, effective stellar velocity dispersions of
$\sigma_\mathrm{e} = 282$ km s$^{-1}$ and $270$ km s$^{-1}$ are
measured for $R_e = $4\farcs7 and 18\farcs3, respectively. We note,
however, that there is just a single measurement of the large-scale
stellar kinematics on each side of the nucleus beyond a radius of
12\arcsec, making it difficult to determine $\sigma_\mathrm{e}$ for a
bulge effective radius of 18\farcs3 from the LRIS data.

These $\sigma_\mathrm{e}$ measurements are based upon the large-scale
kinematics extracted from the major axis LRIS slit, however the
best-fit stellar dynamical model from \S \ref{sec:results} allows us
to predict the luminosity-weighted second moment from a circular
aperture of radius $R_e$. We measured an effective velocity dispersion
of $\sigma_\mathrm{e} = 239$ km s$^{-1}$ when a bulge effective radius
of $R_e = $10\farcs7 is used and the central regions within
$r_\mathrm{sphere}$ are excluded. This method provides a formally more
correct estimate of $\sigma_\mathrm{e}$, as a circular aperture is
being considered instead of relying on a single slit position, but
most of stellar velocity dispersions on the $M_\mathrm{BH} -
\sigma_\star$ relationship were derived using long-slit data and the
definition given by \cite{Gultekin_2009b}. Thus, for consistency
purposes we continue to consider the measurements of
$\sigma_\mathrm{e}$ made from the LRIS data.

With an effective stellar velocity dispersion of $\sigma_\mathrm{e} =
272$ km s$^{-1}$, the black hole mass predicted from the
$M_\mathrm{BH} - \sigma_\star$ relation is $9.4 \times 10^8\ M_\odot$
using the recent calibration of \cite{McConnell_2011b}. Instead, black
hole masses of $6.5 \times 10^8\ M_\odot$ and $4.9 \times 10^8\
M_\odot$ are predicted from the $M_\mathrm{BH} - \sigma_\star$
correlation calibrated by \cite{Graham_2011} and
\cite{Gultekin_2009b}. Thus, our stellar dynamical black hole mass
measurement falls within the expectations of the relationship when
calculating $\sigma_\mathrm{e}$ using the large-scale LRIS kinematics
and a bulge effective radius of 10\farcs7.

\begin{figure*}
\begin{center}
\epsscale{0.7}
\plotone{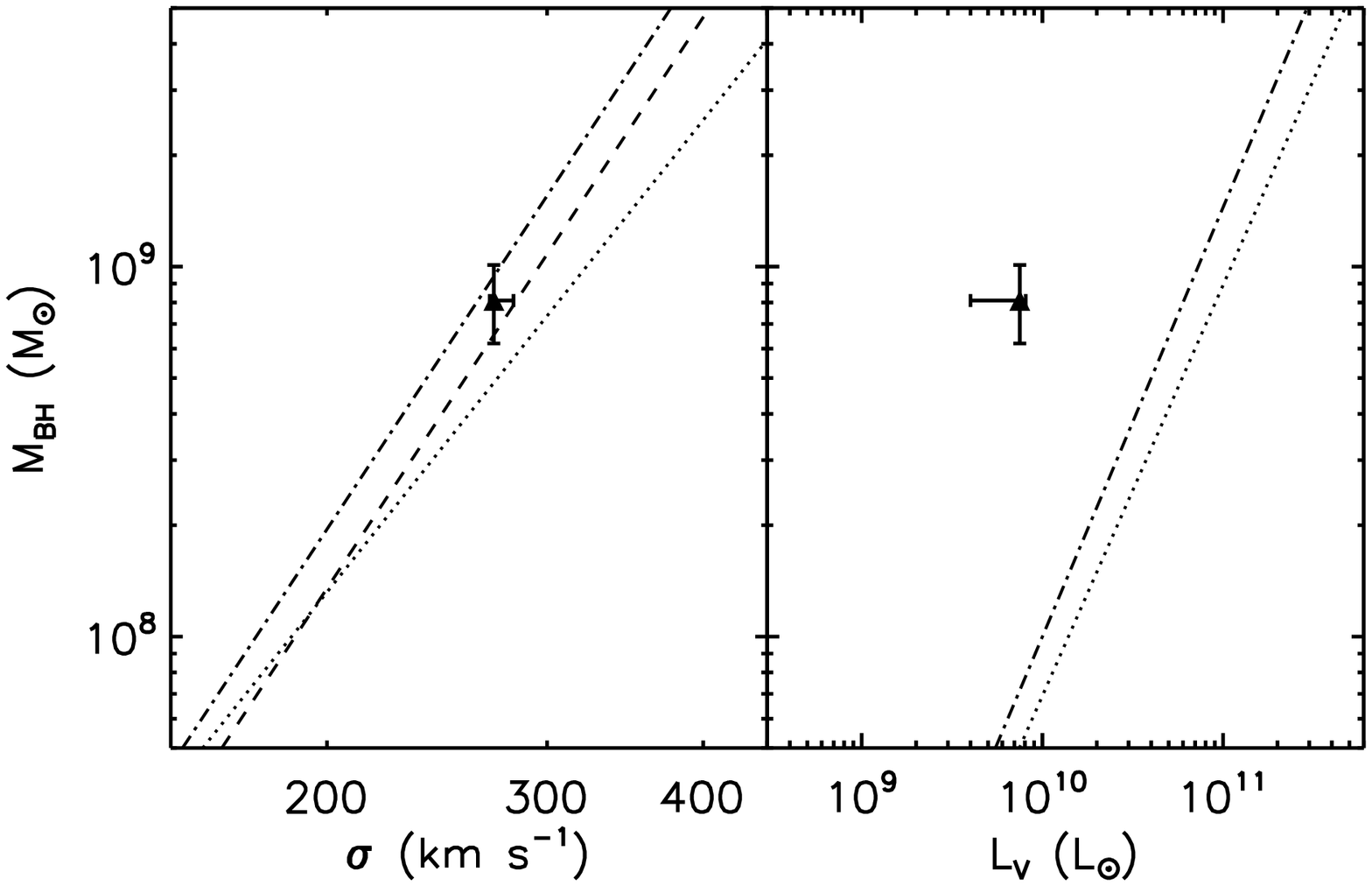}
\caption{Location of NGC 3998 on the $M_\mathrm{BH} - \sigma_\star$
  (left) and $V$-band $M_\mathrm{BH} - L_\mathrm{bul}$ (right)
  relationships. The correlations calibrated by \cite{Gultekin_2009b},
  \cite{Graham_2011}, and \cite{McConnell_2011b} are given by the
  dotted, dashed, and dot-dashed lines, respectively. The errors on
  the bulge stellar velocity dispersion show the range of values that
  were calculated using the large-scale LRIS data between
  $r_\mathrm{sphere}$ and the largest and smallest bulge effective
  radius measurements found in the literature [$R_e = $18\farcs3
  \citep{Baggett_1998} and $R_e = $4\farcs7
  \citep{Sani_2011}]. Similarly, the errors on the bulge luminosity
  were determined by applying the smallest and largest values of $B/T$
  found in the literature [$B/T = 0.41$ \citep{Sani_2011} and $B/T =
  0.83$ \citep{SanchezPortal_2004}] to the total $V$-band luminosity
  given in RC3. \label{fig:n3998_mbhrels}}
\end{center}
\end{figure*}

To place NGC 3998 on the $M_\mathrm{BH} - L_\mathrm{bul}$ correlation,
we use the total, extinction corrected, $V$-band luminosity of $9.7
\times 10^9\ L_\odot$ for NGC 3998 from the Third Reference Catalog of
Bright Galaxies (RC3; \citealt{RC3_catalog}). The average of the
bulge-to-total ratio ($B/T$) values reported by \cite{Fisher_1996} and
\cite{SanchezPortal_2004} is $B/T = 0.77$, which suggests a $V$-band
luminosity of $7.5 \times 10^9\ L_\odot$ for the bulge of NGC
3998. Such a bulge luminosity translates into a prediction of $7.2
\times 10^7\ M_\odot$ and $5.0 \times 10^7\ M_\odot$ for the black
hole mass using the $V$-band $M_\mathrm{BH} - L_\mathrm{bul}$ relation
of \cite{McConnell_2011b} and \cite{Gultekin_2009b}. Even when
deliberately overestimating the bulge luminosity, and using the total,
$V$-band luminosity reported in RC3, the $M_\mathrm{BH} -
L_\mathrm{bul}$ relation suggests a black hole mass of $9.6 \times
10^7\ M_\odot$ \citep{McConnell_2011b}. The stellar dynamical
measurement therefore places NGC 3998 well above the $M_\mathrm{BH} -
L_\mathrm{bul}$ relationship. In Figure \ref{fig:n3998_mbhrels}, we
show the location of NGC 3998 on the black hole mass scaling
relations.

\section{Conclusions}
\label{sec:conclusions}

To conclude, we have studied the stellar dynamics of the nearby, S0
galaxy, NGC 3998. From Keck LGS AO OSIRIS observations in the $K$ band
and archival \emph{HST} STIS data covering the \ion{Ca}{2} triplet
lines, we mapped out the 2D kinematics within $\sim$2\arcsec of the
nucleus with a superb angular resolution of $\sim$0\farcs1, thereby
resolving the gravitational sphere of influence of the black hole. In
addition, we obtained long-slit data at four PAs with Keck LRIS to
measure the large-scale stellar kinematics, extending to $\sim$1
$R_e$, which is essential for constraining the stellar orbital
distribution. We found that the galaxy is rapidly rotating with $V
\sim \pm200$ km s$^{-1}$ and exhibits a very sharp rise in the
velocity dispersion to values of $\sigma \sim$500 km s$^{-1}$. Our
high-quality spectroscopic data further allowed us to quantify the
LOSVD's asymmetric and symmetric deviations from a Gaussian through
the $h_3$ and $h_4$ Gauss-Hermite moments. We observed an
anti-correlation between $h_3$ and $V$, and slight peak in $h_4$ at
the center from the large-scale data.

Combining the kinematics with the luminosity density measured from a
\emph{HST} WFPC2 F791W image and a CFHT WIRCam $K_s$ image, we
constructed three-integral, triaxial, Schwarzschild models. The
intrinsic shape of the galaxy is very round (as round as the imaging
observations allow), and oblate, with axis ratios $p =
0.96_{-0.13}^{+0.04}$ and $q = 0.81_{-0.33}^{+0.00}$ at a radius of
10\farcs7. Although we are unable to place strong constraints on the
shape parameters/viewing orientation, by sampling over them, we were
able to extract a robust measurement of the black hole mass and
$I$-band stellar mass-to-light ratio, finding $M_\mathrm{BH} =
(8.1_{-1.9}^{+2.0}) \times 10^8\ M_\odot$ and $M/L =
5.0_{-0.4}^{+0.3}\ M_\odot/L_\odot$. Additional model grids were run
to assess possible systematic effects. We tested the effect on the
black hole mass when assigning all the light from the central MGE
component to the AGN, incorporating a fixed dark halo model, excluding
the central OSIRIS kinematics that may be biased due to AGN
contamination, adopting two extreme models for the OSIRIS PSF, and
symmetrizing the kinematic measurements. We did not see any
significant changes (outside of the modeling fitting uncertainties) to
the best-fit $M_\mathrm{BH}$ or $M/L$.

With the stellar dynamical black hole mass measurement, NGC 3998 is
consistent with $M_\mathrm{BH} - \sigma_\star$ when using a bulge
stellar velocity dispersion of $272$ km s$^{-1}$, but well off the
$M_\mathrm{BH} - L_\mathrm{bul}$ correlation, even when overestimating
the bulge luminosity. Also, the stellar dynamical black hole mass
measurement is larger than the existing gas dynamical measurement,
with the masses differing by close to a factor of four. NGC 3998 is
now one of eight galaxies for which both stellar and gas dynamical
modeling have been used to measure the mass of the central black
hole. However, the gas kinematics turned out to be strongly disturbed
in two of the galaxies and a clear best-fit mass and the associated
uncertainties could not be measured for another object, preventing a
meaningful comparison between the stellar and gas dynamical
techniques. The five remaining comparison studies have produced mixed
results, ranging from excellent agreement between the two mass
measurements to the stellar dynamical black hole mass exceeding the
gas dynamical determination by a factor of five. Of the three cases in
which the black hole masses disagree, the stellar dynamical
measurement is always larger than the black hole mass derived with gas
dynamical modeling. With such a limited sample, clearly more
cross-checks are necessary before conclusions can be made regarding
the consistency of the gas and stellar dynamical techniques, the
subsequent effects on the $M_\mathrm{BH}$ $-$ host galaxy
relationship, and the magnitude and distribution of the cosmic scatter
in the correlations. Targeting galaxies with dust lanes could be
particularly useful, as well-ordered, symmetric dust lanes suggest the
presence of a regularly rotating gas disk \citep{Ho_2002} and can also
be used to constrain galaxy inclinations.

\acknowledgements

We would like to thank Shelley Wright for her assistance with the
OSIRIS observations, and Ronald L\"{a}sker and his collaborators for
sharing their CFHT WIRCam image of NGC 3998 prior to publication. We
also thank the anonymous referee for a thoughtful and quick response,
which helped improve the manuscript. Research by A.~J.~B. and
J.~L.~W. has been supported by NSF grants AST-0548198 and
AST-1108835. J.~L.~W. has also been supported by an NSF Astronomy and
Astrophysics Postdoctoral Fellowship under Award No. 1102845. Some of
the data presented herein were obtained at the W.~M. Keck Observatory,
which is operated as a scientific partnership among the California
Institute of Technology, the University of California and the National
Aeronautics and Space Administration. The Observatory was made
possible by the generous financial support of the W.~M. Keck
Foundation. We wish to recognize and acknowledge the very significant
cultural role and reverence that the summit of Mauna Kea has always
had within the indigenous Hawaiian community. We are most fortunate to
have the opportunity to conduct observations from this mountain. All
of the stellar dynamical models presented in this work were run on the
THEO cluster at the Rechenzentrum Garching (RZG) of the Max Planck
Society. Some of the data presented in this paper were obtained from
the Multimission Archive at the Space Telescope Science Institute
(MAST). STScI is operated by the Association of Universities for
Research in Astronomy, Inc., under NASA contract NAS5-26555. This
research has made use of the NASA/IPAC ExtragalacticDatabase (NED),
which is operated by the Jet Propulsion Laboratory, California
Institute of Technology, under contract with NASA. We acknowledge the
usage of the HyperLeda database (http://leda.univ-lyon1.fr).

\end{document}